\documentclass{svjour3}                     
\smartqed  

\usepackage{lineno,hyperref}
\usepackage{amsmath}
\usepackage{subfigure}
\usepackage{caption}
\usepackage{xcolor}
\usepackage{amssymb}
\usepackage{lipsum}
\usepackage{amsfonts}
\usepackage{graphicx}
\usepackage{epstopdf}
\usepackage{algorithmic}
\usepackage{amsmath}
\usepackage{bm}
\usepackage{mathrsfs}
\usepackage{amsthm}
\usepackage{soul}
\usepackage{algorithm}
\usepackage{algorithmic}
\usepackage{arydshln}
\usepackage{listings}

\usepackage{listings} 
\usepackage{setspace} 

\definecolor{Code}{rgb}{0,0,0} 
\definecolor{Decorators}{rgb}{0.5,0.5,0.5} 
\definecolor{Numbers}{rgb}{0.5,0,0} 
\definecolor{MatchingBrackets}{rgb}{0.25,0.5,0.5} 
\definecolor{Keywords}{rgb}{0,0,1} 
\definecolor{self}{rgb}{0,0,0} 
\definecolor{Strings}{rgb}{0,0.63,0} 
\definecolor{Comments}{rgb}{0,0.63,1} 
\definecolor{Backquotes}{rgb}{0,0,0} 
\definecolor{Classname}{rgb}{0,0,0} 
\definecolor{FunctionName}{rgb}{0,0,0} 
\definecolor{Operators}{rgb}{0,0,0} 
\definecolor{Background}{rgb}{0.98,0.98,0.98} 

\lstdefinelanguage{Python}{ 
	numbers=left, 
	numberstyle=\footnotesize, 
	numbersep=1em, 
	xleftmargin=1em, 
	framextopmargin=2em, 
	framexbottommargin=2em, 
	showspaces=false, 
	showtabs=false, 
	showstringspaces=false, 
	frame=l, 
	tabsize=4, 
	basicstyle=\ttfamily\small\setstretch{1}, 
	backgroundcolor=\color{Background}, 
	commentstyle=\color{Comments}\slshape, 
	stringstyle=\color{Strings}, 
	morecomment=[s][\color{Strings}]{"""}{"""}, 
	morecomment=[s][\color{Strings}]{'''}{'''}, 
	morekeywords={import,from,class,def,for,while,if,is,in,elif,else,not,and,or,print,break,continue,return,True,False,None,access,as,,del,except,exec,finally,global,import,lambda,pass,print,raise,try,assert}, 
	keywordstyle={\color{Keywords}\bfseries}, 
	morekeywords={[2]@invariant,pylab,numpy,np,scipy}, 
	keywordstyle={[2]\color{Decorators}\slshape}, 
	emph={self}, 
	emphstyle={\color{self}\slshape}, 
}

\begin{document}

\title{Robust design optimisation of continuous flow polymerase chain reaction thermal flow systems
}

\titlerunning{Robust Optimisation of CFPCR}        

\author{Yongxing Wang \and Hazim A. Hamad \and Jochen Voss \and Harvey M. Thompson 
}


\institute{Yongxing Wang \at School of Mechanical Engineering, University of Leeds, Leeds, UK \\
           \email{scsywan@leeds.ac.uk}           
\and
Hazim A. Hamad \at School of Mechanical Engineering, University of Leeds, Leeds, UK \\ \email{mnhsh@leeds.ac.uk} \and
Jochen Voss \at School of Mathematics, University of Leeds, Leeds, UK \\ \email{J.Voss@leeds.ac.uk} \and 
Harvey M. Thompson \at School of Mechanical Engineering, University of Leeds, Leeds, UK \\ \email{H.M.Thompson@leeds.ac.uk}  
}

\date{Received: date / Accepted: date}

\maketitle

\begin{abstract}
This paper presents an efficient methodology for the robust optimisation of Continuous Flow Polymerase Chain Reaction (CFPCR) devices. It enables the effects of uncertainties in device geometry, due to manufacturing tolerances, on the competing objectives of minimising the temperature deviations within the CFPCR thermal zones, together with minimising the pressure drop across the device, to be explored. We first validate that our training data from conjugate heat transfer simulations of the CFPCR thermal flow problems is noise free and then combine a deterministic surrogate model, based on the mean of a Gaussian Process Regression (GPR) simulator, with Polynomial Chaos Expansions (PCE) to propagate the manufacturing uncertainties in the geometry design variables into the optimisation outputs. The resultant probabilistic model is used to solve a series of robust optimisation problems. The influence of the robust problem formulation and constraints on the design conservatism of the robust optima in comparison with the corresponding deterministic cases is explored briefly.

\end{abstract}

\section{Introduction}
\label{sec_introduction}
Precision control of the heat transfer to and from small volumes of liquid flowing in fluidic channels underpins many important practical applications. Examples include environmental pollution monitoring systems, fuel cells and pharmaceutical manufacturing systems \cite{stroock2002chaotic,tarn2018study}. There are also several examples in electronics cooling, where numerous heat sink configurations employ single phase flows in fluidic channels to dissipate high heat fluxes encountered in e.g. radio frequency and microwave applications \cite{agarwal2017modeling}. This paper is motivated by single phase flow in fluidic channels that form part of Continuous Flow Polymerase Chain Reaction (CFPCR) systems used for the rapid amplification of DNA segments. These have played a pivotal role in the public health response to detecting and monitoring COVID-19.

In CFPCR systems the channels are arranged in a serpentine format to perform a rapid thermal cycling, where each straight component incorporates three distinct thermal zones: denaturation at $\sim 95^\circ C$, annealing between $40^\circ C \sim 50^\circ C$ and extension between $60^\circ C \sim 70^\circ C$. Recent studies have demonstrated that CFPCR systems present challenging, multi-objective optimisation problems with key objectives such as the minimisation of total processing time and heating power requirements, and the maximisation of temperature uniformity and DNA amplification efficiency \cite{papadopoulos2015comparison,hamad2021computational}. In addition to the channel geometry, flow speeds, distances between thermal zones, heating arrangements and the thickness, thermal conductivity and biological compatibility of the chip materials all have major influences on the DNA amplification and power consumption.

In these, and indeed all, fluidic channel systems, the geometry of the fluidic channel has a vital influence on thermo-fluid performance. Heat sinks, for example, employ a wide range of vortex generator systems to improve heat transfer and reduce pressure drop \cite{al2018benefits}, while for CFPCR systems, channel sizes, inter-zone spacing and channel cross-sectional shape have all been found to useful variables for manipulating performance \cite{thomas2014thermal}. Channel cross-sectional shape can also be useful variables for manipulating performance. For example, adopting spiral \cite{hashimoto2004rapid} or radial cross-sections \cite{schaerli2009continuous} have been proposed as an effective means of reducing the PCR reaction time,  while \cite{duryodhan2016simple} showed that employing diverging fluidic channels can improve the overall temperature uniformity within the PCR stages, ultimately enhancing DNA amplification efficiency. 

All flow systems using fluidic channels are, however, subject to aleatory uncertainties, due to variations in geometric dimensions or operating conditions, each of which can affect overall performance significantly. Most previous studies have either used deterministic approaches which ignore uncertainties altogether or simply account from them using factors of safety \cite{zhao2016numerical,leng2015multi,zhang2020reliability}. This paper is the first to explore the effect of uncertainties in channel dimensions, arising from manufacturing tolerances, on the performance and optimisation of CFPCR systems.

Techniques for Optimisation Under Uncertainty (OUU) aim to enable designers to produce robust, reliable designs which account for the impact of uncertainties in the input and operating variables on the resultant uncertainties in performance \cite{beyer2007robust}. The two main approaches to OUU are Robust Design Optimisation (RDO) and reliability-based optimisation methods \cite{cavazutti2013optimization}. Reliability-based ones focus on the probability of failure and are particularly important in the design of safety-critical systems, such as in the aerospace industry \cite{shahpar2011challenges}, and in other industries where the consequences of failure can be disastrous, e.g. the financial system \cite{ramadhan2017numerical}. \cite{zhang2020reliability} recently considered the reliability-based optimisation of heat sinks for cooling electronics based on liquid flows through micro-fluidic channels. Using a performance measure approach to reliability, they consider the influence of Gaussian uncertainties in flow rate and heat flux on critical temperature constraints, the violation of which are known to lead to increased component failures. They found that, compared to deterministic designs, higher pumping power is needed under design uncertainties to maintain the same thermal performance. 

Inspired by the pioneering work of \cite{taguchi1986introduction}, RDO methods aim to produce optimal designs that are less sensitive to variable inputs, usually by enforcing a low standard deviation in the output quantities \cite{li2020multidisciplinary}. They generally use optimisation objectives and constraints based on the mean and standard deviation of performance, formulated in a number of ways, either by using objective functions which are specified in terms of weighted sums, compromise or aggregation approaches \cite{lopez2011approximating}. These have been used to account successfully for uncertainties in a range of challenging engineering problems. These include the effects of variations in material properties during sheet metal forming \cite{tang2009robust}, the effect of manufacturing tolerances and variable operating conditions on air-cooled heat sinks \cite{bodla2013optimization}, the influence of manufacturing errors caused by milling and etching on gripper mechanisms and heat sinks \cite{schevenels2011robust}, and in lifecycle assessment methods \cite{wang2014application}. 

Although used widely in the aerospace and automotive industries for many years, the use of optimisation methods employing physics-based flow simulations has expanded rapidly into many other application areas recently \cite{khatir2019cfd}. For deterministic optimisation methods, the key challenges are to develop efficient numerical methods for both flow simulations and design space exploration. When uncertainties are accounted for too, as needed in RDO methods, the computational burden can be increased many-fold since additional sampling is required in order to find information about the Probability Density Function (PDF) of performance objectives (usually the mean and standard deviation) at specific design points \cite{lopez2011approximating}. The use of surrogate modelling of the performance objectives is usually a key component of any feasible optimisation approach--using techniques such as Radial Basis Functions (RBF) \cite{cavazutti2013optimization}, Moving Least Squares (MLS) \cite{gilkeson2013multi} or supervised machine learning methods such as Artificial Neural Networks (ANN) \cite{hamad2021computational} - to create inexpensive surrogates for mapping between input design variables and output objectives which can be used within optimisation algorithms \cite{haftka2016parallel}. For RDO the PDF of objectives can be generated using simple approaches such as Monte Carlo and Latin Hypercube sampling, but these may require thousands of numerical simulations in order to generate the statistical information and are often simply unfeasible. 

In this study we will explore the effects of aleatoric uncertainties in the fluidic channel geometry, caused by manufacturing tolerances, on the design and optimisation of CFPCR systems. We will use an efficient approach to RDO by combining surrogate modelling based on Gaussian Process Regression (GPR) and Polynomial Chaos Expansion (PCE) to develop the required statistical information around design points. GPR is an efficient probabilistic machine learning approach \cite{williams2006gaussian} that can learn the mean and probability density function of outputs and has been shown to provide effective statistical surrogates for many problems \cite{hopfe2012robust,domingo2020using,gao2018developing,nhu2019advanced} and is therefore well-suited to RDO. We will alleviate the additional cost of multiple flow simulations, needed to generate statistical moments of the objectives, using generalised polynomial chaos expansion, which provides a rigorous approach to propagating the uncertainties in input design variables into the output variables of interest \cite{yang2017polynomial,feinberg2015chaospy}. PCE provides much better convergence rates than random sampling \cite{bodla2013optimization}, and has been used successfully in numerous RDO studies -- see e.g. \cite{bodla2013optimization} RDO applied to heat sink design. 

The paper is organised as follows. Section \ref{cfd_data} describes the propotype CFPCR system being investigated, the specification of the conjugate heat transfer problem, the numerical methods for its solution and the validation of the methods. Section \ref{sec_gpr_pce} presents the formulation of the RDO problems, together with the surrogate modelling and PCE methods used for generating the statistical information used within the RDO. Section \ref{sec_implementation} contains a series of results which explore the effect of uncertainty in fluidic channel geometry on CFPCR performance and compare the results of the RDO with those from a corresponding deterministic approach. Conclusions are drawn in section \ref{sec_conclusion}. 

\section{Numerical methods}\label{cfd_data}
The thermal flow problem considered is based on the prototypical CFPCR flow analysed in our recent paper \cite{hamad2021computational}, so only brief details are given here.

\subsection{Problem description}
\begin{figure}[h!]
	\centering  
	\includegraphics[width=3.5in,angle=0]{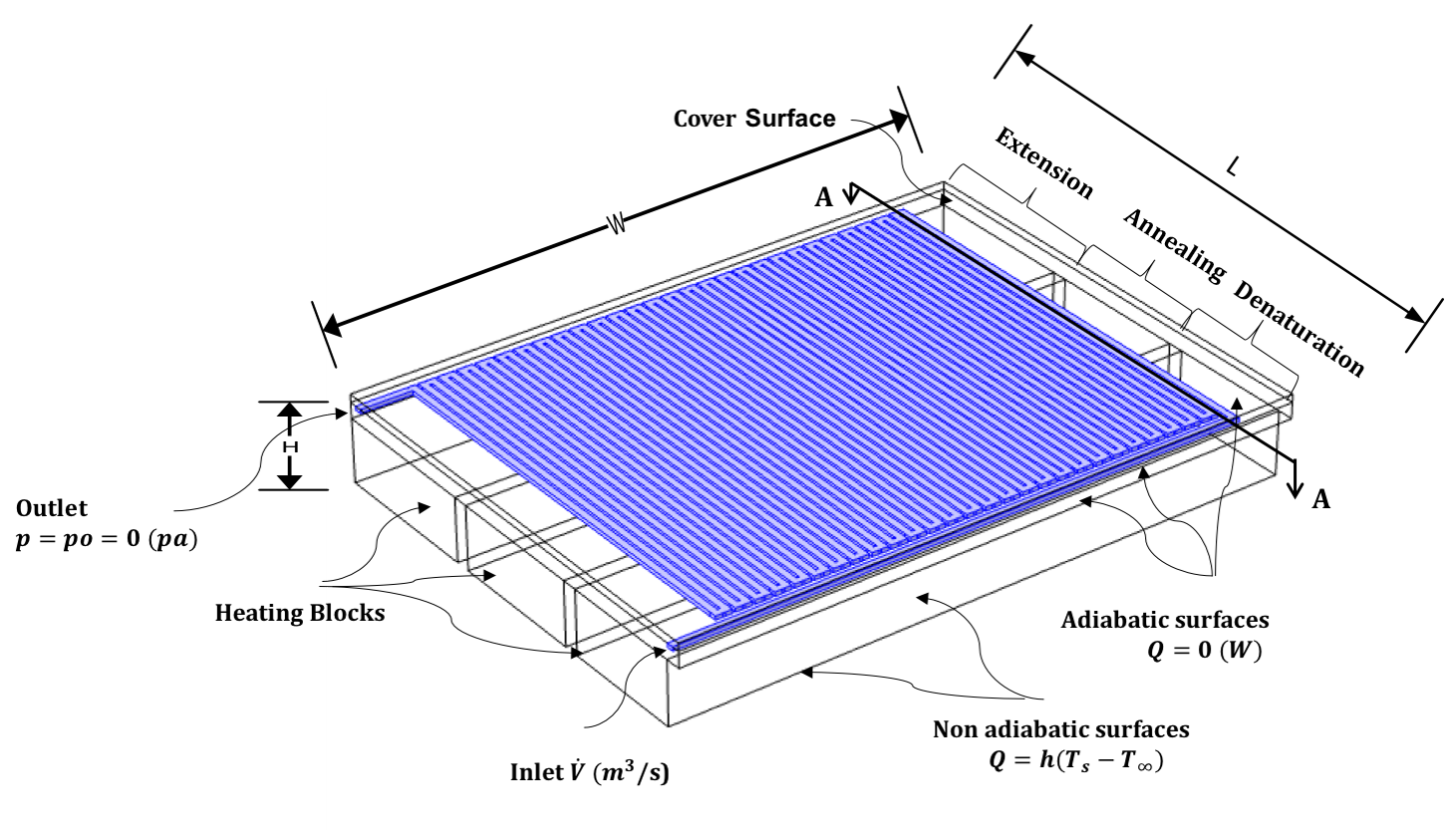}
	\captionsetup{justification=centering}
	\caption {\scriptsize A schematic diagram of the serpentine microfluidic channel and heating arrangements and the hydrodynamic and thermal boundary conditions.} 
	\label{Picture1}
\end{figure}

\begin{figure}[h!]
	\centering  
	\includegraphics[width=2.5in,angle=-0]{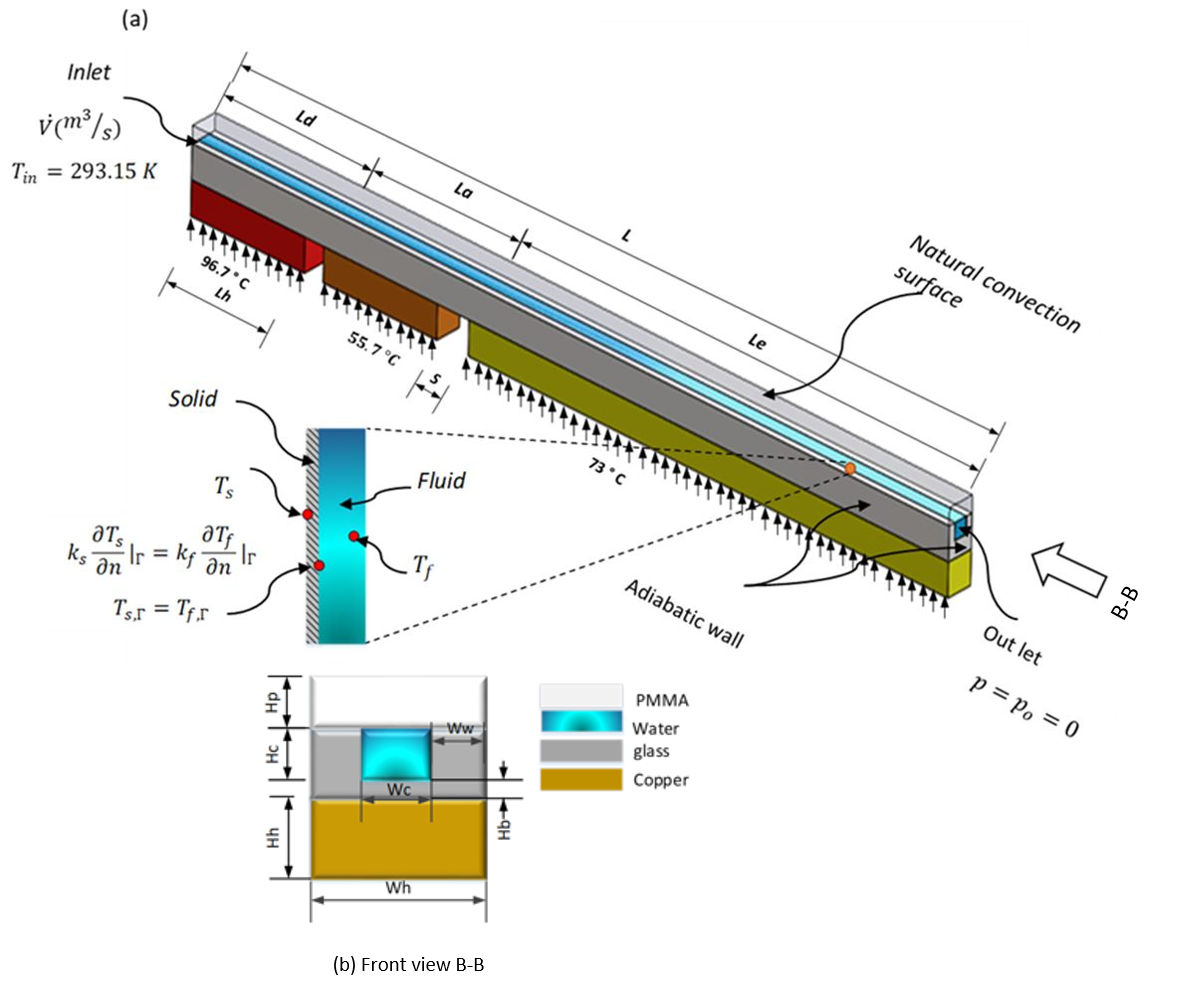}
	\captionsetup{justification=centering}
	\caption {\scriptsize Schematic diagram of a prototype of PCR channel representing a section of the serpentine microfluidic channel.} 
	\label{Picture2}
\end{figure}

Continuous Flow Polymerase Chain Reaction (CFPCR) systems are typically based on a serpentine fluidic channel arrangement, see Figure \ref{Picture1}, in order to create a thermal cycling procedure which amplifies DNA segments, allowing detection and identification of gene sequences. Within each straight channel component there are three thermal stages of denaturation (~$95^oC$), annealing (~$56^oC$) and extension (~$72^oC$) and these are represented by the prototypical CFPCR thermal flow problem shown in Figure \ref{Picture2}.

The fluidic geometry with a glass substrate and PMMA cover material, where $W_c$, $H_c$, $W_w$, $H_b$ and $L$ are the channel width, height, wall thickness (the spacing between the channels), the bottom height and total length respectively. Three individual heaters are placed under the glass chip with a constant separation $S=1mm$, and the length of the denaturation, annealing and extension zones are chosen to give residence time ratios of $1:1:4$ respectively.                                                                                                                                        

\subsection{Conjugate heat transfer modelling}
We adopt a conjugate heat transfer model of a steady, single-phase, laminar flow employed by previous studies \cite{moschou2014all,chen2008temperature,aziz2016numerical,chiu2017small}. The liquid is water, with temperature dependent density, thermal conductivity and viscosity. The effects of radiation and buoyancy are neglected and there is no internal heat source. The governing equations include the Navier-Stokes equations 
\begin{equation}
	\rho_f(T)\left({\bf u}\cdot\nabla\right){\bf u}
	=\nabla\cdot\left[\mu_f\left(T_f \right)\left(\nabla{\bf u}+(\nabla{\bf u})^T\right)-p{\bf I}-\frac{2}{3}\mu_f\left(T_f \right)(\nabla\cdot{\bf u}){\bf I}\right],
\end{equation}
\begin{equation}
	\nabla\cdot\left(\rho_f{\bf u}\right)=0,
\end{equation}
with ${\bf u}$ and $p$ being the fluid velocity and pressure respectively. The governing equations also include the heat transfer equations in the fluid
\begin{equation}\label{heat_equation}
	\rho_f(T)C_{f}\left(T\right){\bf u}\cdot\nabla T_f
	=\nabla\cdot\left(k_f(T)\nabla T_f\right),
\end{equation}
and in the solid
\begin{equation}
	\nabla\cdot\left(k_s\nabla T_s\right)=0,
\end{equation}
where $C_{f}$, $k_f$ and $k_s$ represent the specific heat and thermal conductivities of the fluid and solid respectively. The thermo-physical properties $\rho_f{(T)}$, $\mu_f{(T)}$, $C_{p}(T)$ and $k_f(T)$ depend on the temperature as follows \cite{bergman2011fundamentals}:
\begin{eqnarray}
	&&\rho_f(T)=838.466+1.4T-0.003T^2+3.72\times10^{-7}T^3, \\
	&&\mu_f(T)=1.38-0.02T-1.36\times10^{-4}T^2-4.64\times10^{-7}T^3 +8.9\times10^{-10}T^4,\\
	&& C_{f}(T)=12010.15-80.41T+0.31T^2-5.38\times10^{-4}T^3+3.63\times10^{-7}T^4,\\
	&& k_{f}(T)=-0.869+0.00895T-1.584\times10^{-4}T^2-7.975\times10^{-9}T^3.
\end{eqnarray}
The thermal conductivity of copper is $k_s=400 W/(m\cdot K)$ and the target temperatures in the denaturation, annealing and extension zones are $95^\circ C$, $55^\circ C$ and $72^\circ C$ respectively. The governing equations are solved using COMSOL Multiphysics 5.4, see \cite{hamad2021computational} for further details.
\subsection{Performance metrics}
The metrics used in the optimisation are the temperature uniformity in the zones and the pressure drop along the channel. High temperature uniformity leads to high levels of DNA amplification within the PCR process, while minimising the latter reduces both the structural stresses in the chip and the hydraulic power input needed to pump the liquid through the chip. Temperature uniformity is quantified in terms of deviations from the target temperature {$T_{target}$} in the zones by the following discrete $L^2$-error in the corresponding zone:
\begin{equation}
	T_{dev}=\|T_{f}-T_{target}\|,
\end{equation}
where $\|\cdot\|=\frac{1}{N}\sqrt{\sum_{k=1}^{N}(\cdot)^2}$, with $N$ being the number of discrete points of our numerical experiments. The pressure drop along the channel is simply
\begin{equation}
	\Delta p=p_{in}-p_{out}.
\end{equation}
\subsection{Numerical validation}
The mesh convergence is examined by obtaining numerical solutions on a series of structured finite element grids of increasing refinement, on a desktop PC with Microsoft Windows 10 and 32GB physical RAM. Table 1 shows the number of Degrees Of Freedom (DOF), physical memory (PM), Virtual Memory (VM), execution time and calculated $T_{dev}$ and $\Delta p$ for each mesh. There is only a small change ($<0.2\%$) in the obtained results if the number of mesh is above 877552. The numerical results presented  below have been obtained on the mesh with 877552 elements as an appropriate compromise between computational expense and accuracy.

\begin{table}[h!]
	\small
	\centering
	\begin{tabular}{ |c|c|c|c|c|c|c|}
		\hline
		NO. of elements & DOF ($\times 10^5$) & PM (GB) & VM (GB) & Time (s) & $\Delta p (Pa)$ & $T_{dev} (^\circ C)$ \\
		\hline
		93024 &	2.52 &	2.71 &	3.28 &	119	& 51.31	&  27.98 \\
		\hdashline
		216096 &	5.1679 &	3.17 &	3.92 &	303	&  51.00 &	27.81  \\
		\hdashline
		380480 &	8.3090 &	3.89 &	4.67 &	553	&  50.64  &  27.70 \\
		\hdashline
		448686 &	14.485 &	4.47 &	5.26  &	677	&  50.58  &	27.58 \\
		\hdashline
		877552 &	17.763 &	6.56 &	7.42 &	1839 &	50.48 &	27.57 \\
		\hdashline
		1026602 &	20.373 &	8.23 &	9.52 &	3849 &	50.45 &	27.50 \\
		\hdashline
		1171632	& 23.329 &	7.93 &	8.92 &	4050 &	50.43 &	27.52 \\
		\hline
	\end{tabular}
	\captionsetup{justification=centering}
	\caption {Effect of mesh density for the case of $H_c=W_c=500\mu m$.}
	\label{Effect_of_grid}
\end{table}

\begin{figure}[h!]
	\begin{minipage}[t]{0.5\linewidth}
		\centering  
		\includegraphics[width=2.4in,angle=0]{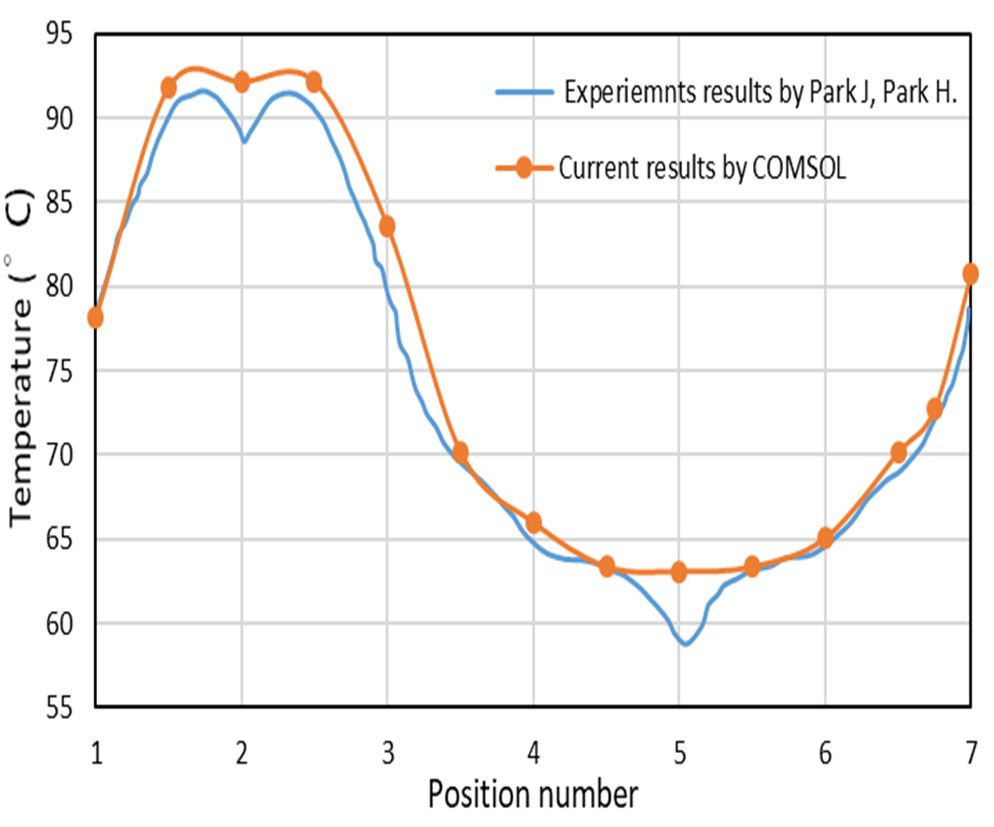}	    
		\captionsetup{justification=centering}		
		\caption {\scriptsize Comparison of surface temperature variation along flow direction in a diverging microchannel.} 
		\label{comparison_of_surface_temperature}
	\end{minipage}
	\begin{minipage}[t]{0.5\linewidth}
		\centering  
		\includegraphics[width=2.5in,angle=0]{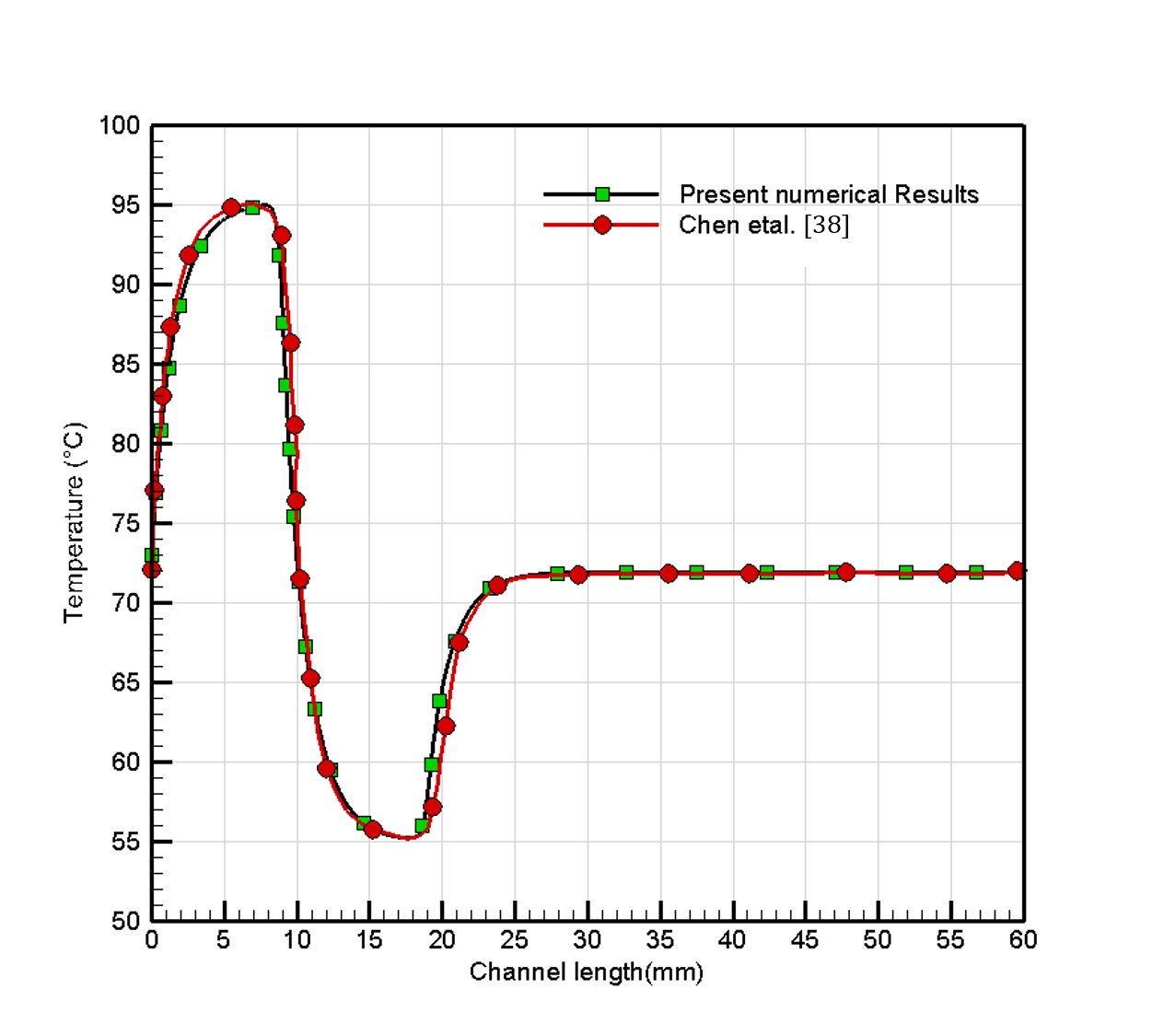}	   
		\captionsetup{justification=centering} 		
		\caption {\scriptsize Comparison of fluid average temperature profile along the centreline.} 
		\label{comparison_of_fluid_average}
	\end{minipage}     		
\end{figure}

The numerical model is first compared with the experimental results in \cite{park2017thermal} for thermal flow in diverging channels. Figure \ref{comparison_of_surface_temperature} shows a generally very good agreement between the experimental and numerical results. The next comparison is with the numerical results in \cite{chen2008temperature} where $H_c=150 \mu m$ and $W_c=50 \mu m$. The numerical predictions of the temperature profile along the three temperature zones, shown in Figure \ref{comparison_of_fluid_average}, are also in very good agreement with the published results.

\section{Gaussian process regression and polynomial chaos expansions}\label{sec_gpr_pce}
We briefly introduce the Gaussian Process Regression (GPR) and Polynomial Chaos Expansion (PCE) methods used to generate the statistical information used within the RDO.
\subsection{Gaussian process regression}\label{sec_gpr}
One Conjugate Heat Transfer (CHT) simulation provides one data point: $\left(W_c, H_c,\Delta p, T_{dev}\right)$. We generate simulation data at $n=100$ input points $\left(W_c, H_c\right)\in[0.015, 0.5]\times[0.05, 0.15]$, including 20 evenly spaced points at boundaries and 80 random, uniformly distributed points inside the domain. These generate the corresponding output variables $\left(\Delta p, T_{dev}\right)$ which are found to lie within $[50.87, 1437]\times[12.87, 16.29]$. We deliberately use data at the boundaries to improve the accuracy of extrapolation outside of the domain, which is required when applying the PCE method, as discussed in the following section \ref{sec_pce}.

For notation's convenience, let $\left(x_{1},x_{2},y_{1},y_{2}\right)_j$ ($j=1,2,\ldots,n$) denote the $n$ data points from the CHT simulations with input ${\bf x}=\left(x_{1},x_{2}\right)$ and corresponding output $\left(y_{1},y_{2}\right)$. The GPR can learn the relation $f_i(\cdot)$ between ${\bf x}$ and $y_i$ ($i=1,2$) from these $n$ training data points, i.e. it computes a surrogate model $f_i({\bf x})\sim\mathcal{N}\left(\mu_{f_i}^{gpr},(\sigma_{f_i}^{gpr})^2\right)$ with the superscript $``gpr"$ denoting the mean or standard deviation from the GPR model:
\begin{equation}\label{output_mu}
	\mu_{f_i({\bf x})}^{gpr}={\bf k}^T\left({\bf K}+\alpha{\bf I}\right)^{-1}{\bf y}_i,
\end{equation}
and
\begin{equation}\label{output_sigma}
	\sigma_{f_i({\bf x})}^{gpr}=k({\bf x},{\bf x})-{\bf k}^T\left({\bf K}+\alpha{\bf I}\right)^{-1}{\bf k}.
\end{equation}
In the above, ${\bf y}_i=\left(y_{i1},y_{i2},\ldots,y_{in}\right)^T$ ($i=1,2$) is a column vector from the observations of the $i^{th}$ component of the output variables, and $\alpha$ is a specified noise in this observation ${\bf y}_i$. $k=k\left({\bf x}_p,{\bf x}_q\right)=\sigma_f^2exp\left(-\frac{1}{2l^2}\left|{\bf x}_p-{\bf x}_q\right|^2\right)$ is the squared-exponential covariance function, with $\sigma_f$ and $l$ being hyperparameters, which can be determined by cross validation \cite{williams2006gaussian,mcgibbon2016osprey} or maximising the marginal likelihood \cite{williams2006gaussian}, and
\begin{equation}
	{\bf k}=\left[k({\bf x}, {\bf x}_1), k({\bf x}, {\bf x}_2), \ldots, k({\bf x}, {\bf x}_n)\right]^T,
\end{equation}

\begin{equation}
	{\bf K} = \begin{bmatrix} 
		k({\bf x}_1, {\bf x}_1) & k({\bf x}_1, {\bf x}_2) & \dots \\
		\vdots & \ddots & \\
		k({\bf x}_n, {\bf x}_1) &        & k({\bf x}_n, {\bf x}_n)
	\end{bmatrix}.
\end{equation}

The output standard deviation in (\ref{output_sigma}) of the GPR model can be interpreted as uncertainties from two different sources: one is the noise $\alpha$ of our input training data $(y_1, y_2)$, and the other one is the discrete error of our training data themselves, due to the finite dataset. It is reasonable to assume that data points from CHT simulations are clean and noise free \cite{sacks1989design,simpson2008design}, so that we will be able to specify a very small $\alpha$ when training our GPR model in Section \ref{noise_free_surface}. For the second type of uncertainty, we shall demonstrate in Section \ref{sec_probabilistic} (see Figure \ref{sigma_ratio}) that it is much smaller than the manufacturing errors considered in this paper. We can therefore use the GPR method to create a deterministic surrogate model for analysing the propagation of manufacturing errors. 
\subsection{Polynomial chaos expansions}\label{sec_pce}
Real manufacturing processes create inevitably a certain level of noise in the input geometrical parameters. Our approach is to assume a reasonable control error in the input parameters and use the PCE method to propagate this noise to the output variables. These resultant probabilistic surrogate models are then used to solve optimisation problems in Section (\ref{sec_optimisation}).

Considering an error $e_x$ around an input point ${\bf x}=(x_1,x_2)$, we need to know the corresponding error $e_y$ in the outputs $y_i=f_i({\bf x})$ ($i=1,2$). A fast analysis of this problem is to use the Taylor expansion to express $y_i$ around ${\bf x}$ if the derivative of $f_i$ is available. Alternatively, a convenient approach might be to create random points (based upon a distribution assumption such as $e_x\sim\mathcal{N}(0,\sigma_n^2)$) around the input ${\bf x}$, and calculate the corresponding statistics around the output $y_i=f_i({\bf x})$. The former is an efficient approach, but unfortunately we cannot easily access the derivative of $f_i$ for our problem. The latter Monte Carlo method is very slow (see Appendix \ref{appendix_pce_convergence} for two tests using this method).  

The PCE method is much more efficient than the Monte Carlo approach (see \ref{appendix_pce_convergence}), and only needs a few data points around ${\bf x}$ in order to calculate the error at the output -- this is available at any point in the design space using our deterministic GPR surrogate model discussed in Section \ref{sec_gpr}. The PCE method approximates the output $y_i$ ($i=1,2$) as a linear combination of orthogonal polynomial basis $\varphi_k({\bf x})=\varphi_{k_1}(x_1)\varphi_{k_2}(x_2)$,
\begin{equation}\label{pce_expression_truncated}
	y_i\approx \sum_{k=0}^m\beta_k\varphi_k({\bf x}).
\end{equation}
Both $x_i$ and $y_i$ ($i=1,2$) are regarded as random variables. We assume our input variables satisfy Gaussian distribution, which requires Hermite PCE basis functions. Other distributions require different basis functions \cite{feinberg2015chaospy}.

There are intrusive and non-intrusive PCE methods to compute the PCE coefficients $\beta_k$ $(k=0,\ldots,m)$. The former requires modification of the governing equations of the system under study, while the latter are sampling-based methods requiring solutions of the governing equations for specific values of the random variables considered \cite{sudret2015polynomial}. Non-intrusive methods provide data-driven models based on experimental or simulation data. Here, we use a non-intrusive method, for which there are pseudo-spectral projection and linear regression methods to compute the coefficients based on design variable sampling. Both of these methods are implemented in ChaosPy \cite{feinberg2015chaospy}, which is used in this paper.

\section{Implementation}\label{sec_implementation}
\begin{figure}[h!]
	\centering  
	\includegraphics[width=2.3in,angle=0]{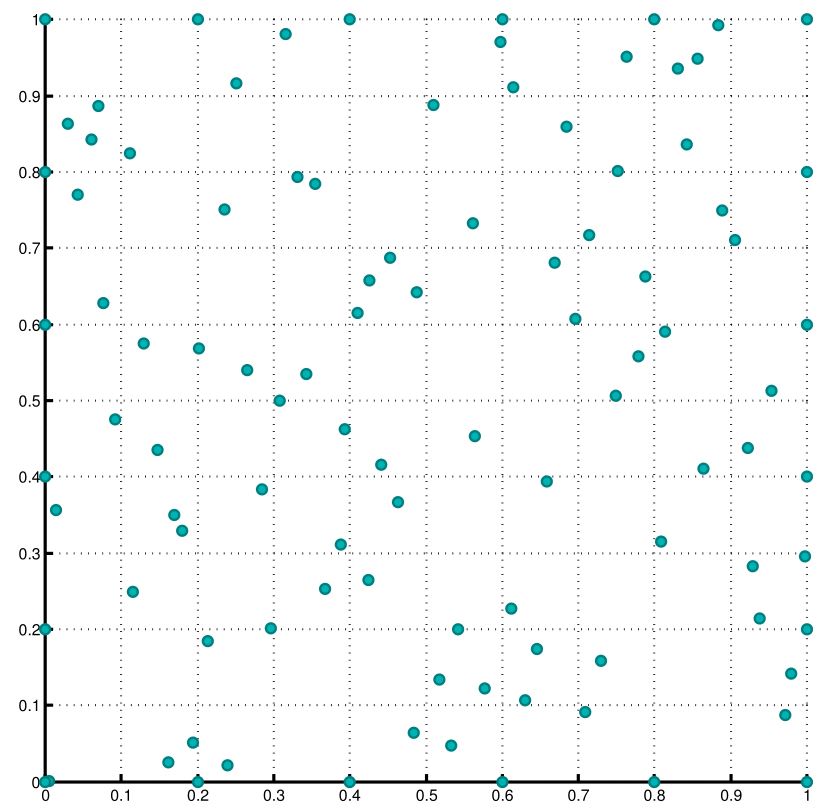}
	\captionsetup{justification=centering}
	\caption {\scriptsize Training data points: 20 evenly spaced points on boundaries and 80 random, uniformly distributed points inside the domain.} 
	\label{training_points}
\end{figure}

The software packages used to implement the GPR method is {\bf GPy} \cite{hensman2012gpy}, the software package to implement the PCE method is {\bf ChaosPy} \cite{feinberg2015chaospy}, and the software package to implement the optimisation problem is {\bf SciPy} \cite{beyreuther2010obspy}. All the Python codes are available in Appendices \ref{appendix_pce_convergence} to \ref{appendix_pce_optimisation}. For presentational convenience, all the design variable training points and outputs $\left(x_{1},x_{2},y_{1},y_{2}\right)_j$ ($j=1,2,\ldots,n$) are normalised to the range $[0,1]$ by
\begin{equation}\label{normaisation}
	\frac{x_i-x_i^{min}}{x_i^{max}-x_i^{min}}, \quad
	\frac{y_i-y_i^{min}}{y_i^{max}-y_i^{min}}, \quad
	i=1,2
\end{equation}
before feeding the training data to the GPR or PCE codes. We have $n=100$ training data points as shown in Figure \ref{training_points}. The results can be easily transformed from normalised space to physical space using (\ref{normaisation}). The corresponding scales in the physical domain are given when necessary.
We also generate a test dataset of $N_{test}=100\times 100$ evenly spaced points in the domain $\Omega=[-0.1,1.1]\times[-0.1,1.1]$ in order to test the surrogate model. The domain is extended by $0.1$ around the boundary of the original domain $\Omega_0=[0,1]\times[0,1]$, so that we can extrapolate values to compute the error propagation at the boundaries using the PCE method as discussed in Section \ref{sec_pce}.
\subsection{Deterministic surrogate model} \label{noise_free_surface}
\begin{figure}[h!]
	\centering  
	\includegraphics[width=5in,angle=0]{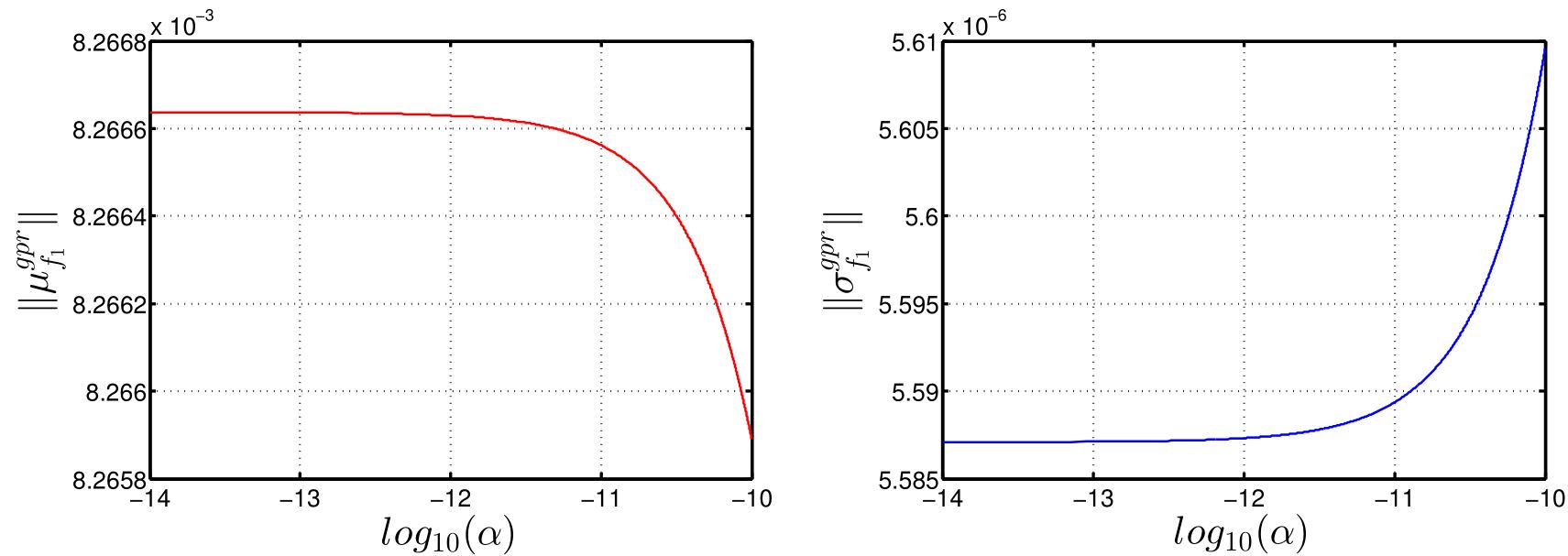}
	\includegraphics[width=5in,angle=0]{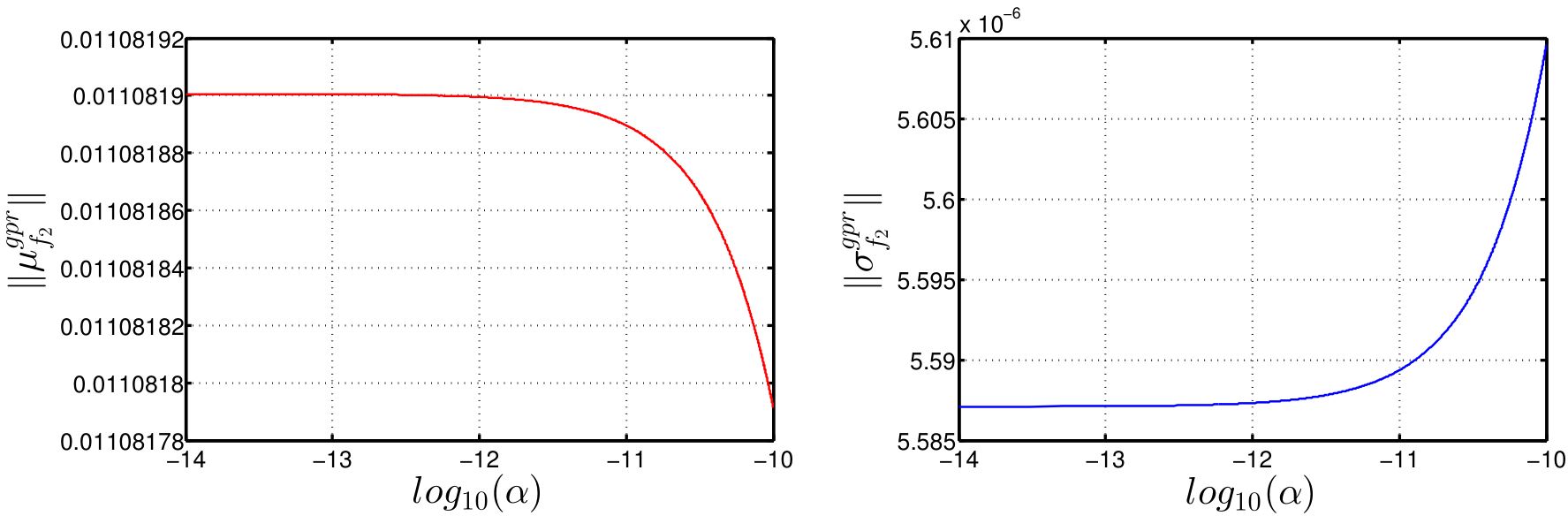}  	
	\captionsetup{justification=centering}
	\caption {\scriptsize Mean and standard deviation for $f_1$ (top) and $f_2$ (bottom) as a function $\alpha$, where $\|\cdot\|=\frac{1}{N_{test}}\sqrt{\sum_{k=1}^{N_{test}}(\cdot)^2}$.} 
	\label{mean_sigma_alpha}
\end{figure}

\begin{figure}[h!]
	\centering  
	\includegraphics[width=2.3in,angle=0]{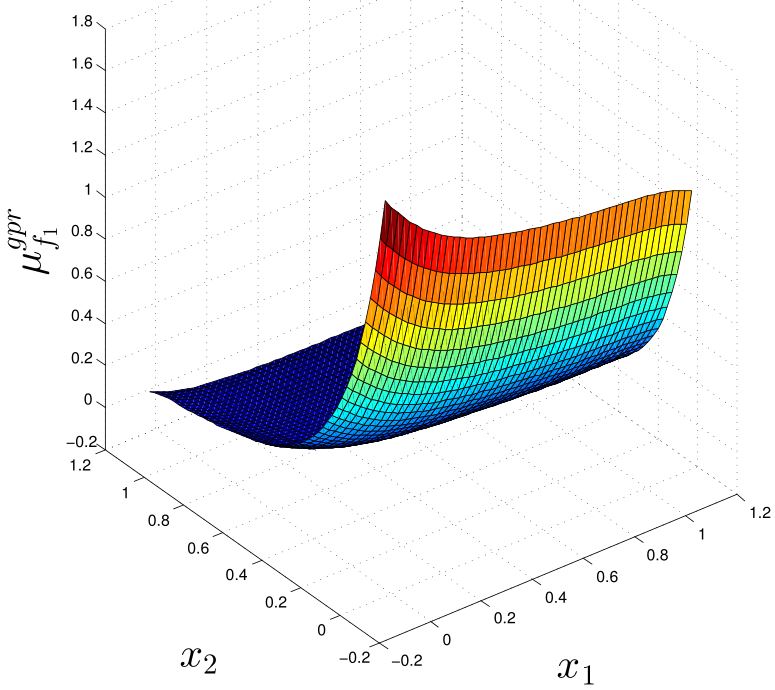}
	\includegraphics[width=2.1in,angle=0]{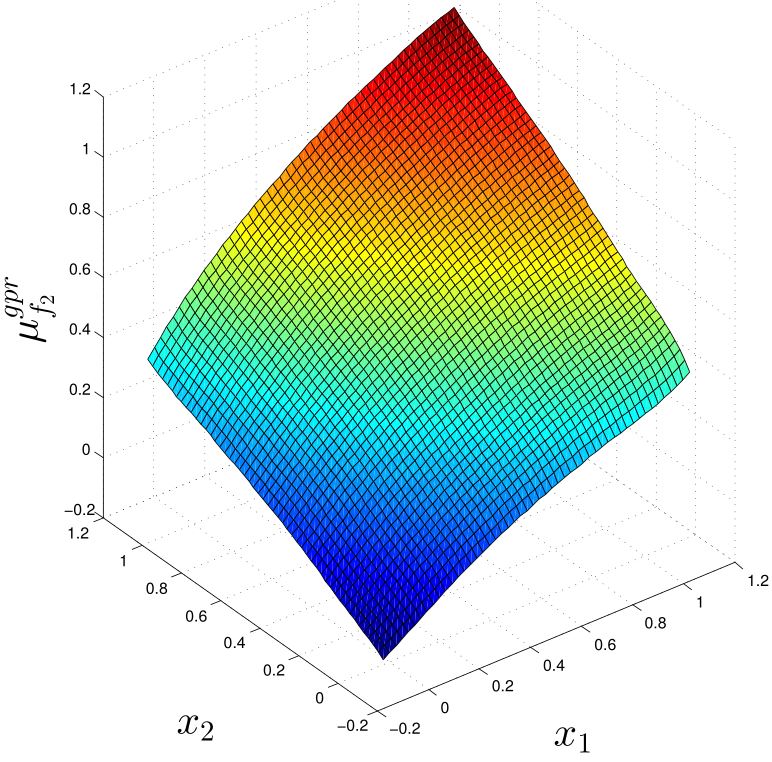}
	\includegraphics[width=2.3in,angle=0]{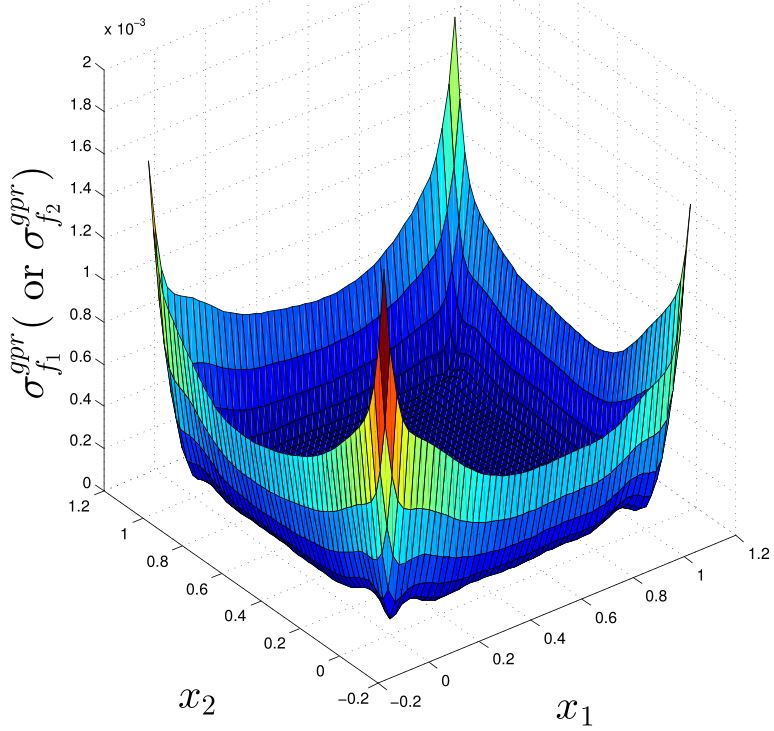}
	\includegraphics[width=2.1in,angle=0]{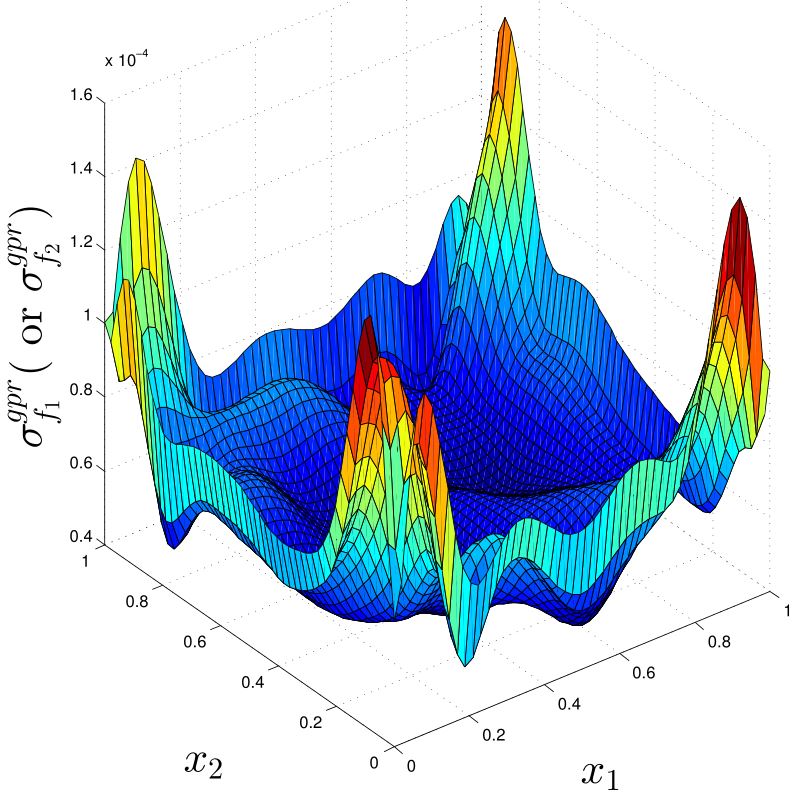}
	\captionsetup{justification=centering}
	\caption {\scriptsize The mean (top) of $f_1({\bf x})$ and $f_2({\bf x})$, and standard deviation plotted on $\Omega$ (left-bottom) and $\Omega_0$ (right-bottom).} 
	\label{response_surface}
\end{figure}

As discussed in Section \ref{sec_gpr} one can specify a small noise parameter $\alpha$ for data from the CHT simulations \cite{sacks1989design,simpson2008design}. Here, we investigate the effect of varying $\alpha$ from $10^{-14}$ to $10^{-10}$ on the accuracy and stability of the GRP algorithm by comparing the predicted mean and standard deviation. We plot the predicted norm of the mean and standard deviation as a function $\alpha$ in Figure \ref{mean_sigma_alpha}, from which it can be seen that the mean effectively constant, and there is no instability issue when using such small values of $\alpha$. Therefore, we will use $\alpha=10^{-12}$ in the following to create the deterministic GPR model. The mean and standard deviation response surfaces are plotted in Figure \ref{response_surface}. Notice from (\ref{output_sigma}) that the standard deviation only depends on the input data, so we have $\sigma_{f_1}^{gpr}=\sigma_{f_2}^{gpr}$. We will also show, in Section \ref{sec_probabilistic}, that this standard deviation $\sigma_{f_1}^{gpr}$ (or $\sigma_{f_2}^{gpr}$) from the GRP model is negligible compared with the noise induced from the manufacturing errors.

\subsection{Probabilistic surrogate model}\label{sec_probabilistic}
We assume a manufacturing error $e_x=0.05$ in the input training data set, so that we have 95\% confidence that an input data point $x_i$ ($i=1,2$) lies in $[x_i-e_x, x_i+e_x]$. Under the assumption of $e_x\sim\mathcal{N}(0,\sigma_n^2)$, we have $\sigma_n=e_x/2=0.025$. We then use the PCE method to propagate this error to the outputs and create a probabilistic surrogate model, using the deterministic GPR surrogate model to calculate function values around point ${\bf x}=(x_1, x_2)$. However, before doing this we have to answer two questions:
\begin{itemize}
	\item what is the appropriate order for the polynomial basis used in the PCE method?
	\item is the standard deviation (uncertainty from discrete data) of the GPR surrogate model small enough, so that we can use the mean $\mu_{f_i}^{gpr}$ (as a deterministic model) to calculate function values?
\end{itemize}
These two questions will be answered in the following section \ref{sec_pce_convergence}.
\subsubsection{Convergence of the PCE method}\label{sec_pce_convergence}
In order to compute the PCE coefficients at a point ${\bf x}\in\Omega_0$, we need several quadrature points $\tilde{\bf x}$ around ${\bf x}$ (pseudo-spectral projection method \cite{feinberg2015chaospy}). The input of these quadrature points are determined by ${\bf x}$ and the order of the polynomial basis of the PCE method, and the outputs are computed by our GPR surrogate model. However, instead of using $\mu_{f_i(\tilde{\bf x})}^{gpr}$ to directly compute the ouptut values, we should consider the uncertainty $\sigma_{f_i(\tilde{\bf x})}^{gpr}$ as well; we cannot neglect $\sigma_{f_i(\tilde{\bf x})}^{gpr}$ by directly comparing its magnitude with the manufacturing error $\sigma_n=e_x/2$, because $\sigma_n$ is from the input space while $\sigma_{f_i(\tilde{\bf x})}^{gpr}$ is in the output space.

We test the PCE method at several different points inside the domain $\Omega_0$ as well as on its boundaries. In order to test the influence of $\sigma_{f_i}^{gpr}$ on the error propagation, we add Gaussian noise, generated by $\mathcal{N}\left(0,\left(\sigma_{f_i(\tilde{\bf x})}^{gpr}\right)^2\right)$, to $\mu_{f_i(\tilde{\bf x})}^{gpr}$ for every point $\tilde{\bf x}$ to compute the PCE coefficients as discussed in Section \ref{sec_pce} -- the Python implementation is given in Appendix \ref{appendix_pce_convergence}.

\begin{figure}[h!]
	\begin{minipage}[t]{0.5\linewidth}
		\centering  
		\includegraphics[width=2.0in,angle=0]{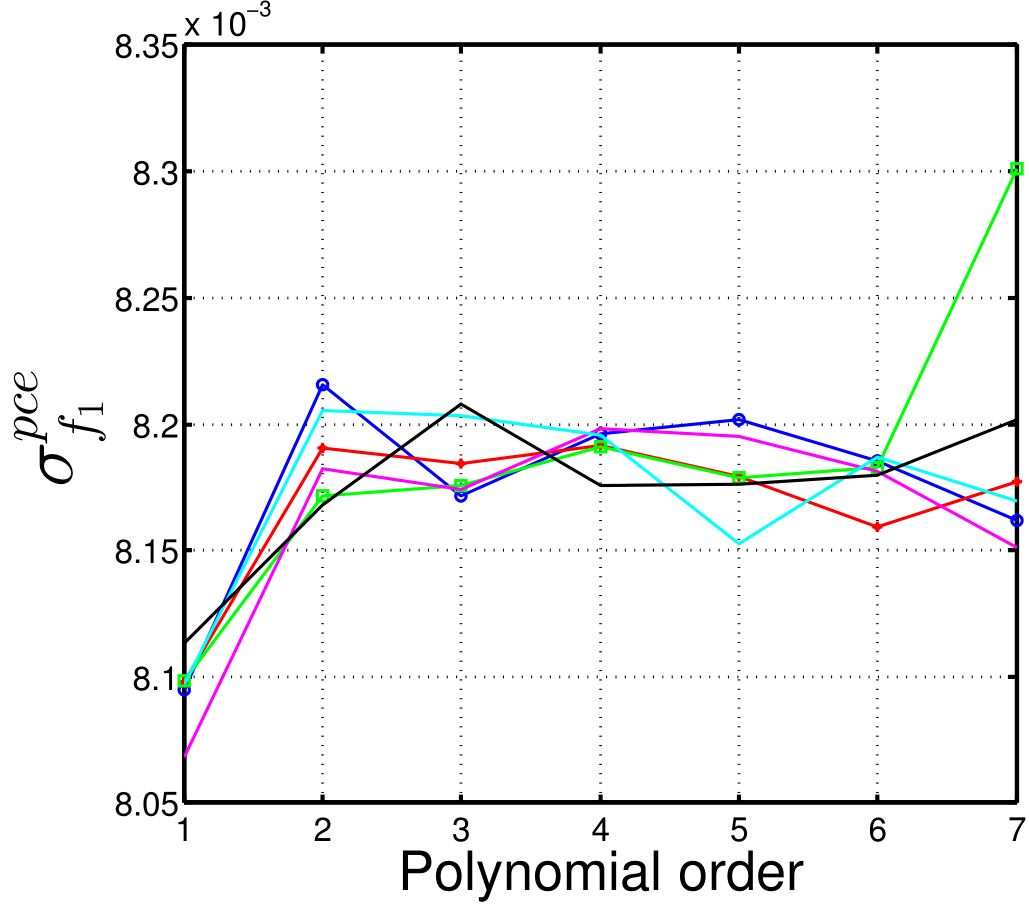}
		\captionsetup{justification=centering}
		\caption*{\scriptsize(a) At point $(0.5, 0.5)$ and $\sigma_{f_1}^{pgr}(0.5, 0.5)=4.9814\times10^{-5}$.}
	\end{minipage}
	\begin{minipage}[t]{0.5\linewidth}
		\centering  
		\includegraphics[width=2.2in,angle=0]{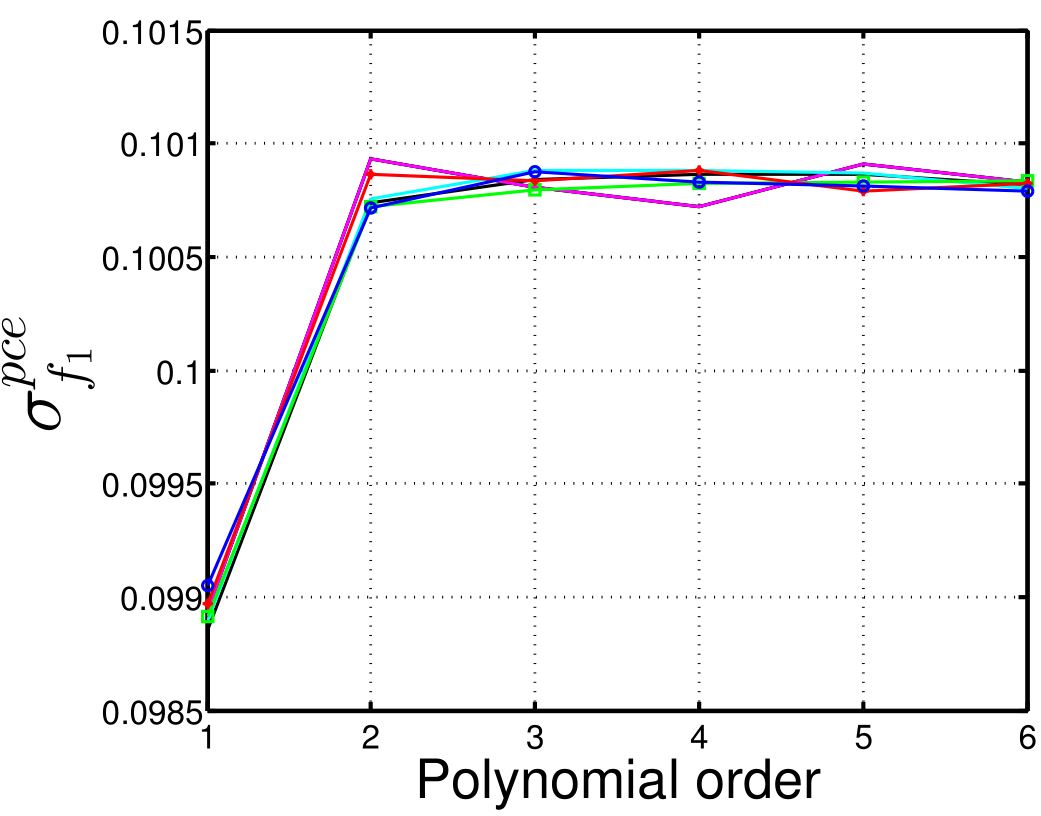}
		\captionsetup{justification=centering}
		\caption*{\scriptsize(b) At point $(1, 0)$ and $\sigma_{f_1}^{pgr}(1, 0)=9.9812\times10^{-5}$.}
	\end{minipage}
	\begin{minipage}[t]{0.5\linewidth}
		\centering  
		\includegraphics[width=2.5in,angle=0]{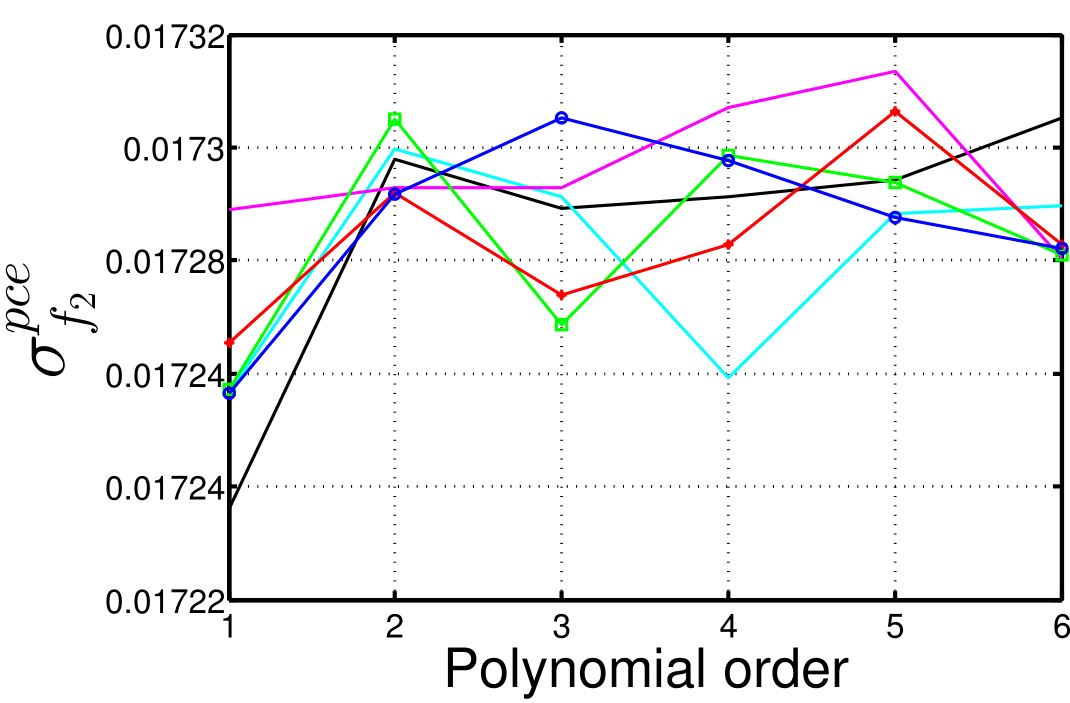}
		\captionsetup{justification=centering}
		\caption*{\scriptsize(c) At point $(0.5, 0.5)$ and $\sigma_{f_2}^{pgr}(0.5, 0.5)=4.9814\times10^{-5}$.}
	\end{minipage}
	\begin{minipage}[t]{0.5\linewidth}
		\centering  
		\includegraphics[width=2.1in,angle=0]{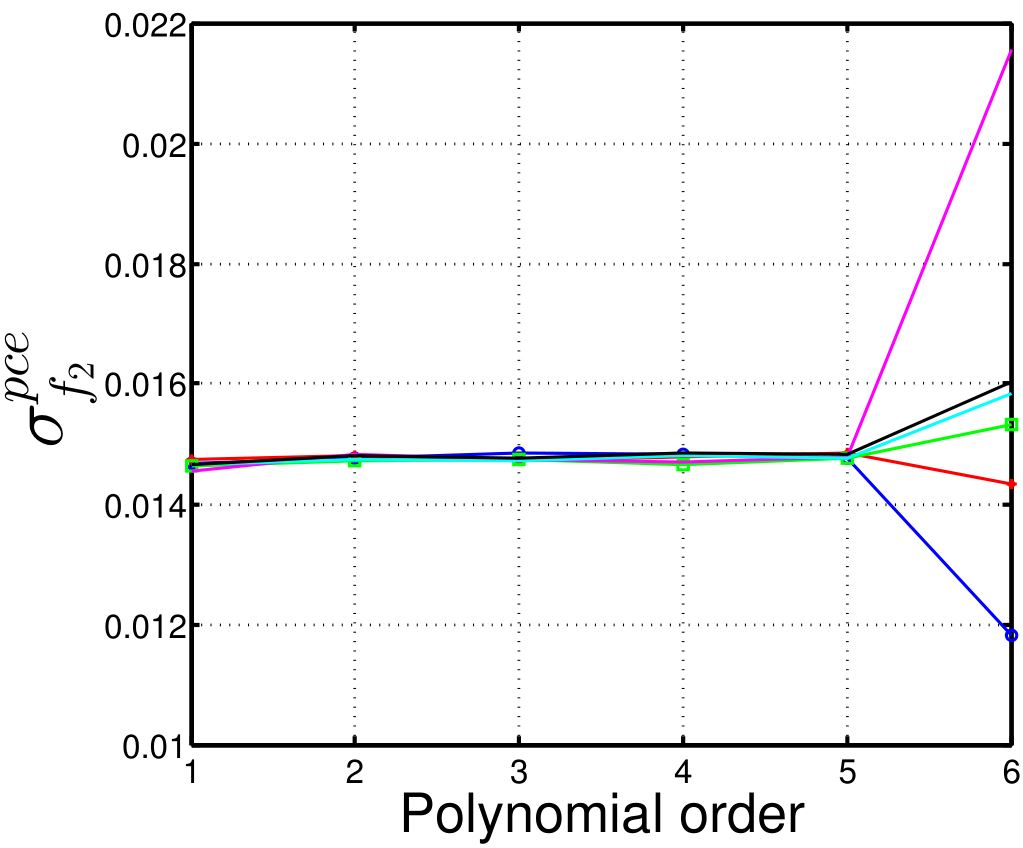}
		\captionsetup{justification=centering}
		\caption*{\scriptsize(d) At point $(1, 1)$ and $\sigma_{f_2}^{pgr}(1, 1)=9.9805\times10^{-5}$.}
	\end{minipage}
	\captionsetup{justification=centering}
	\caption {\scriptsize Error propagation using the PCE method at different points for six cases of random noise in each figures.} 
	\label{pce_convergence}
\end{figure}

\begin{figure}[h!]
	\centering  
	\includegraphics[width=2.3in,angle=0]{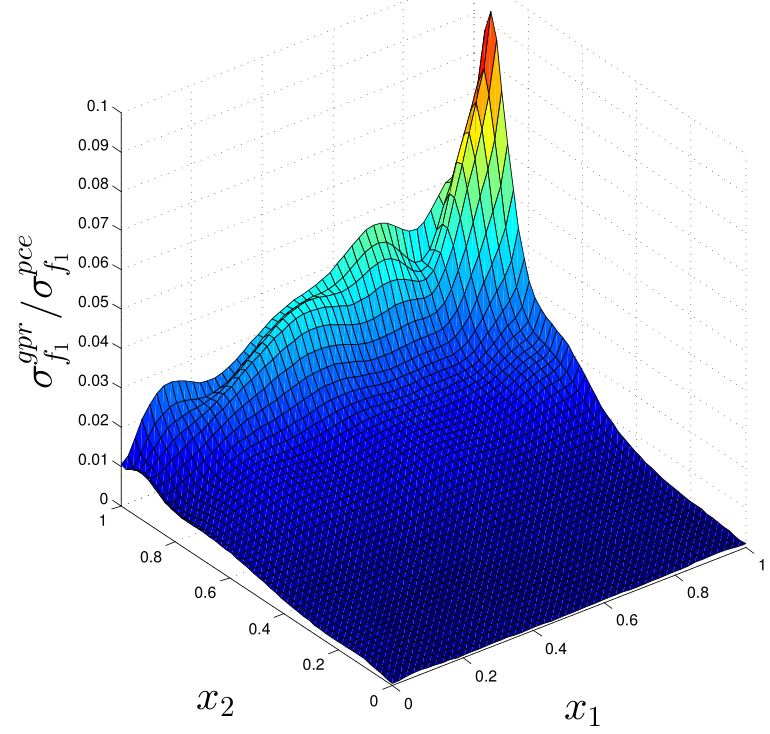}
	\includegraphics[width=2.3in,angle=0]{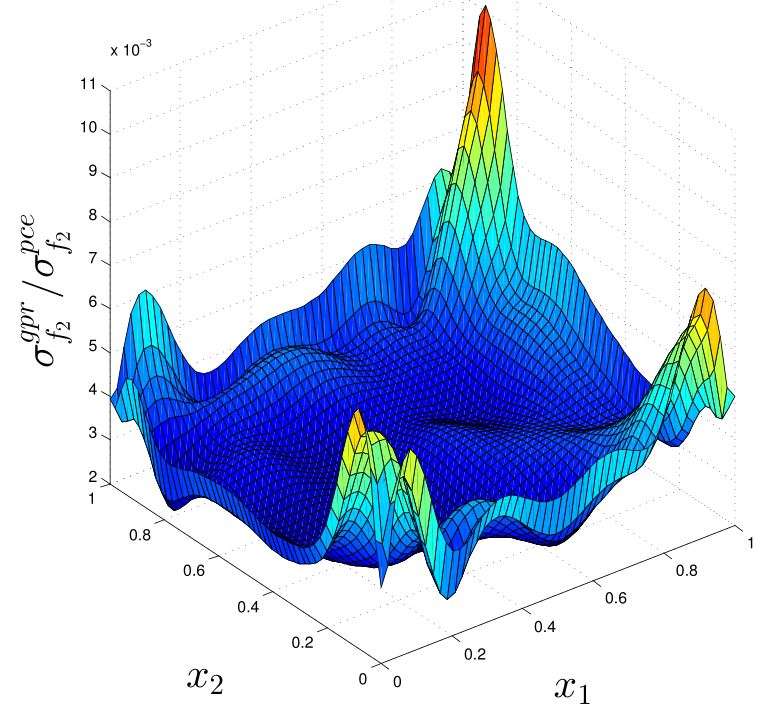}
	\captionsetup{justification=centering}
	\caption {\scriptsize The ratio of the standard deviation of the GPR model to the standard deviation of the PCE model.} 
	\label{sigma_ratio}
\end{figure}

We report in Figure \ref{pce_convergence} the error at three particular points for six cases of random noise, from which it can be seen that a $3^{rd}$ order polynomial basis would be sufficiently accurate to approximate the error propagation.  Notice that in Figure \ref{pce_convergence} the superscript $``pce"$ denotes the mean or standard deviation computed using the PCE method, corresponding to the superscript ``$gpr$" (used through this paper) for the GPR model. In order to compute the coefficients of this $3^{rd}$ order polynomial (using the pseudo-spectral projection method \cite{feinberg2015chaospy}), we need to evaluate $\mu_{f_i}^{gpr}$ at points $0.05836$ away from the boundary of $\Omega_0$, which can be achieved by the extrapolation of our GPR model. Also notice that higher order PCE polynomial basis is needed to evaluate $\mu_{f_i}^{gpr}$ at points further away from the boundary. This introduces extrapolation errors as well, and this can be observed from Figure \ref{pce_convergence} (a) and (d) using a $7^{th}$ and $6^{th}$ order of polynomials respectively. We also present the corresponding standard deviation of the GPR model in the captions in Figure \ref{pce_convergence}, from which we can see this standard deviation of the GPR model is negligible compared with the standard deviation induced by the manufacturing errors. This observation is further validated in Figure \ref{sigma_ratio} where the ratio of the standard deviation of the GPR model to the standard deviation of the PCE model (where the random noise is considered as described above) is less than $0.1$ for $f_1$ and $0.001$ for $f_2$. Although not shown here, this is found to be the case for all examples considered.
In order to further demonstrate that the uncertainty of the GRP model is negligible, we compare the case of adding noise $\mathcal{N}\left(0,\left(\sigma_{f_i(\tilde{\bf x})}^{gpr}\right)^2\right)$ to $\mu_{f_i(\tilde{\bf x})}^{gpr}$ and the case which directly uses $\mu_{f_i(\tilde{\bf x})}^{gpr}$ without noise. Let $\mu_{f_i}^{pce_n}$ ($\sigma_{f_i}^{pce_n}$) and $\mu_{f_i}^{pce}$ ($\sigma_{f_i}^{pce}$) denote the corresponding mean (standard deviation) of the PCE model with and without noise respectively. We find that both $\|\mu_{f_i}^{pce}-\mu_{f_i}^{pce_n}\|$ and  $\|\sigma_{f_i}^{pce}-\sigma_{f_i}^{pce_n}\|$ $\left(\text{with }\|\cdot\|=\frac{1}{N_{test}}\sqrt{\sum_{k=1}^{N_{test}}(\cdot)^2}\right)$ are negligibly small. For a specific random noise, we have $\|\mu_{f_1}^{pce}-\mu_{f_1}^{pce_n}\|=5.7263\times 10^{-7}$, $\|\mu_{f_2}^{pce}-\mu_{f_2}^{pce_n}\|=5.6831\times 10^{-7}$, $\|\sigma_{f_1}^{pce}-\sigma_{f_1}^{pce_n}\|=4.5739\times 10^{-7}$ and $\|\sigma_{f_2}^{pce}-\sigma_{f_2}^{pce_n}\|=4.5521\times 10^{-7}$. Therefore, we shall neglect the uncertainty of the GPR model and use its mean as a deterministic surrogate model in the following sections.
\subsubsection{Noise propagation and response surface with confidence region}\label{noise_surface}
\begin{figure}[h!]
	\centering  
	\includegraphics[width=2.3in,angle=0]{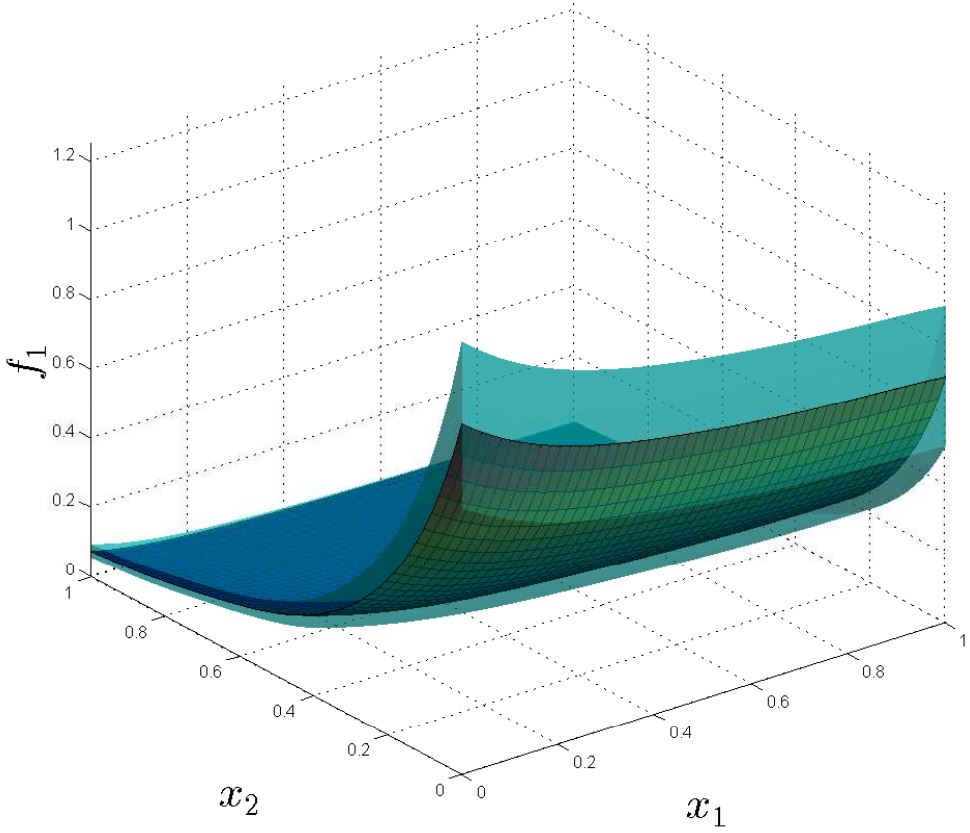}
	\includegraphics[width=2.3in,angle=0]{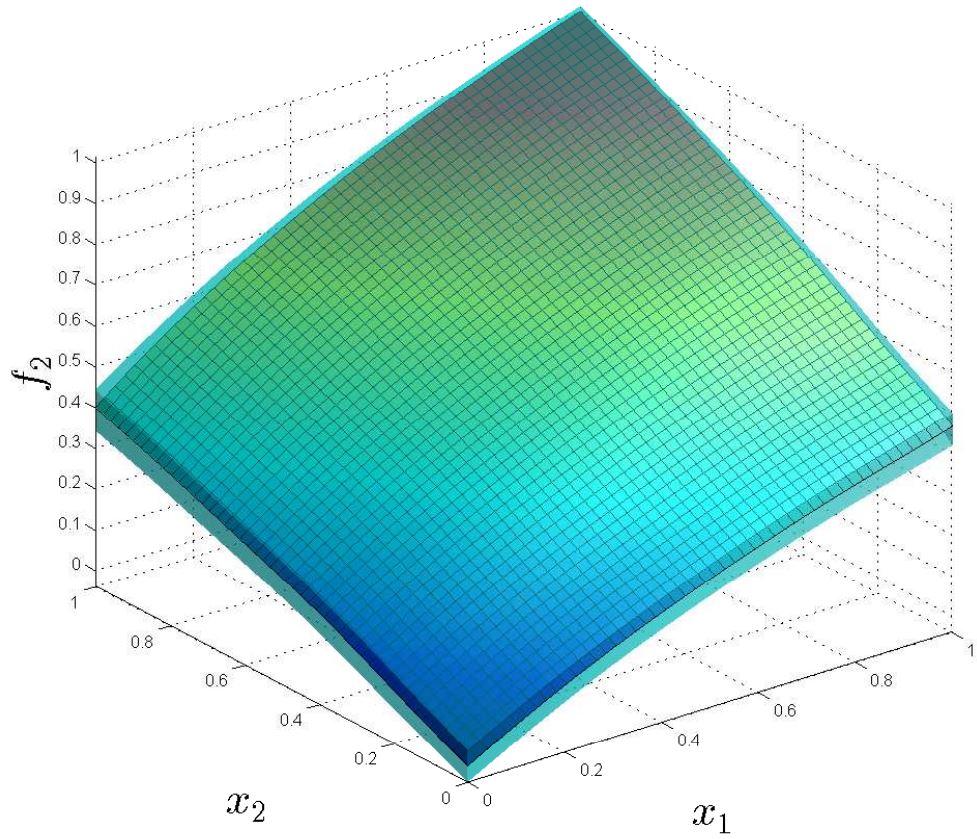}
	\captionsetup{justification=centering}
	\caption {\scriptsize Response surfaces with 95\% confidence region using the PCE method to propagate an error of $5\%$ (of the maximum: 1 in the normalised space, and $0.5$ for $W_c$ and $0.15$ for $H_c$ in the physical space) from the input to the output.} 
	\label{response_surface_confidence}
\end{figure}

\begin{figure}[h!]
	\centering  
	\includegraphics[width=3in,angle=0]{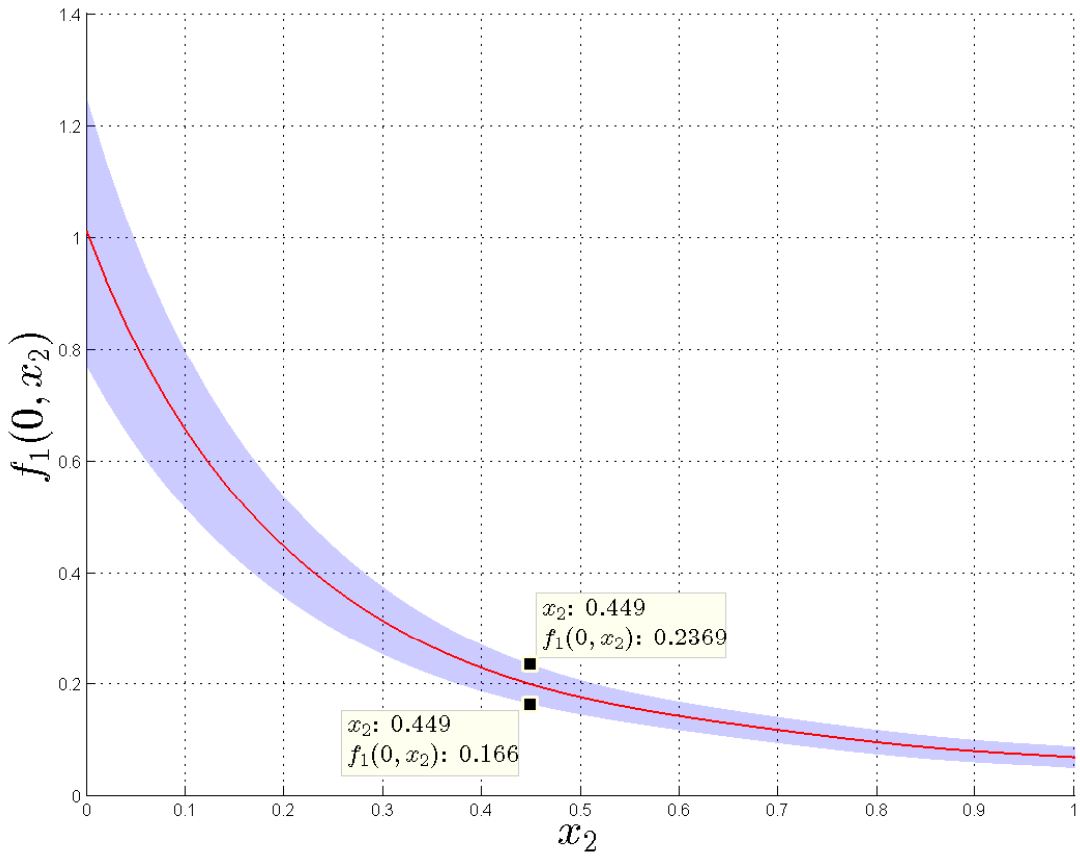}
	\includegraphics[width=3in,angle=0]{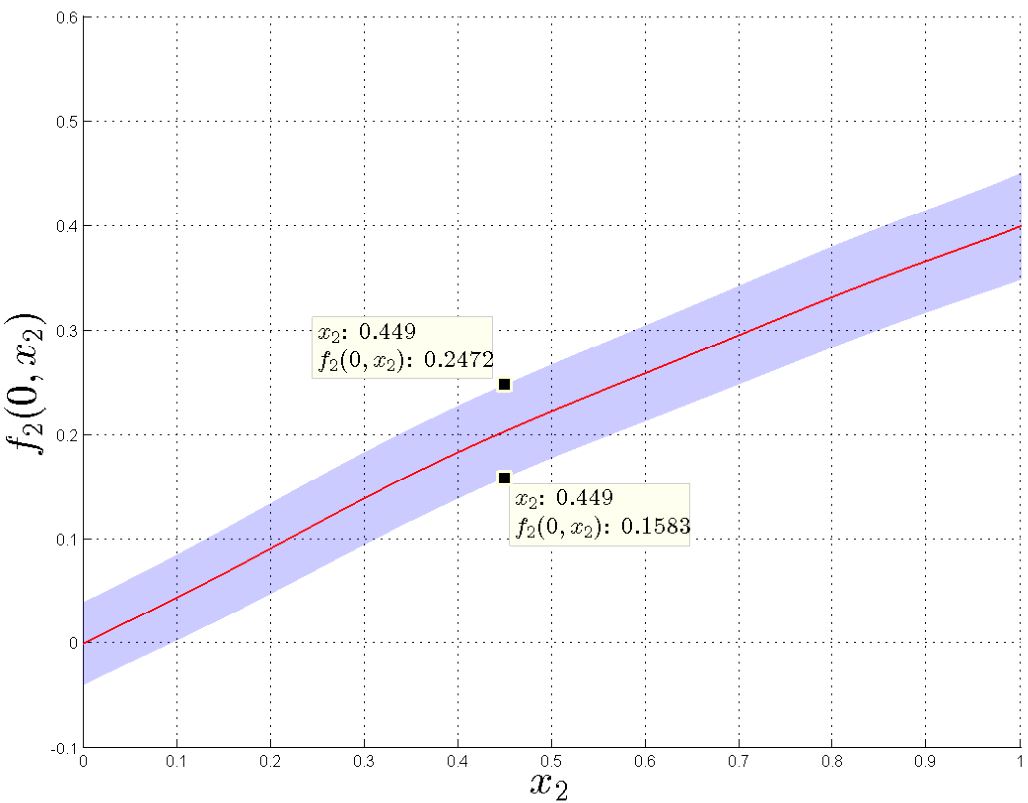}
	\captionsetup{justification=centering}
	\caption {\scriptsize Response surfaces projected to $x_1=0$ with 95\% confidence interval.} 
	\label{response_surface_confidence_projection}
\end{figure}

We can now use the $3^{rd}$ order polynomial, the mean of the GPR model, and propagate the input errors to the outputs and create a probabilistic surrogate model, which incorporates the uncertainty from manufacturing process. The response surfaces with a $95\%$ (two standard deviation) confidence region are plotted in Figure \ref{response_surface_confidence}, from which it can be seen that the input error is amplified where the response surface is steep. This is consistent with analysis using Taylor expansions. We also notice that the second design variable $x_2$ has less influence on objective $f_1$ as shown in Figure \ref{response_surface_confidence}: $f_1(x_1,x_2)$ is almost constant for the same $x_2$, and the two design variables have almost equal influences on objective $f_2$: $f_2(x_1,x_2)$ is visually symmetric along $x_1=x_2$. We plot in Figure \ref{response_surface_confidence_projection} the projection of this mean surface of $f_1(x_1,x_2)$ to $x_1=0$ with a confidence interval for increased clarity.

\subsection{Robust optimisation}\label{sec_optimisation}
\begin{figure}[h!]
\begin{minipage}[t]{0.5\linewidth}
\centering  
\includegraphics[width=2.3in,angle=0]{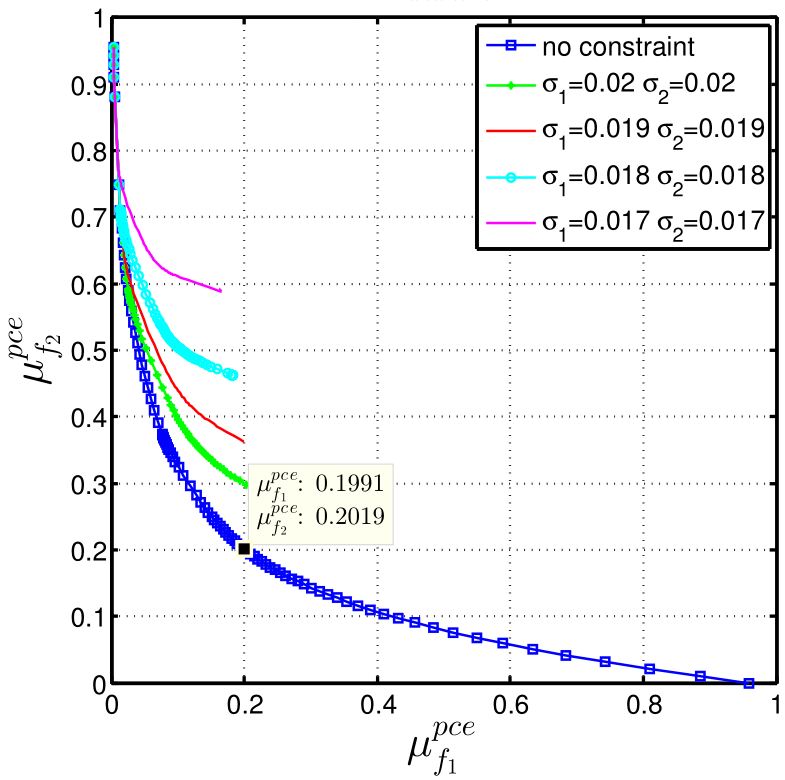}
\captionsetup{justification=centering}
\caption*{\scriptsize(a) Pareto curve.}
\end{minipage}
\begin{minipage}[t]{0.5\linewidth}
\centering  
\includegraphics[width=2.3in,angle=0]{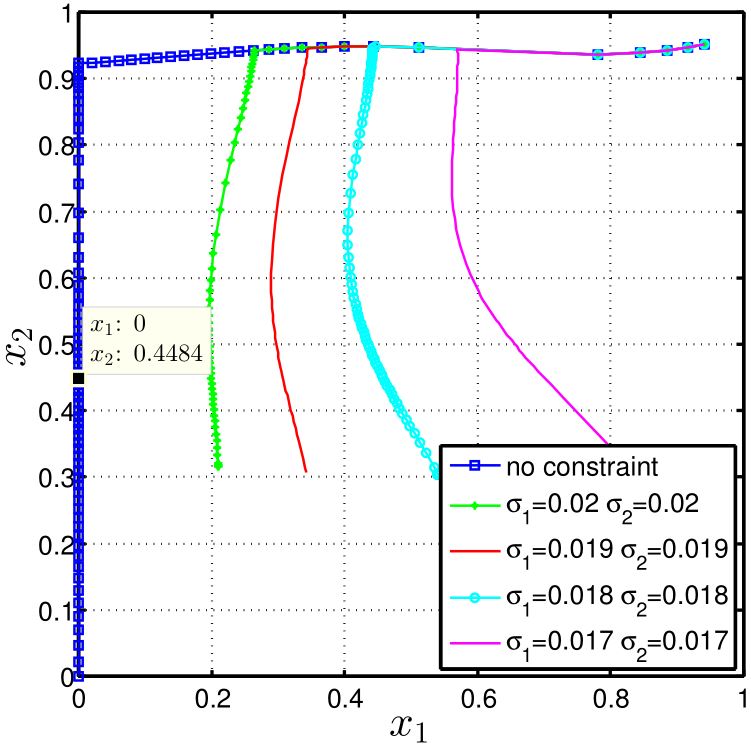}
\captionsetup{justification=centering}
\caption*{\scriptsize(b) Opimal design space.}
\end{minipage}
\caption {\scriptsize Pareto curve  ${\bf p(\omega)}$ and the corresponding designs for different cases of constraints.} 
\label{pareto_constraint_1}
\end{figure}

The probabilistic surrogate model is now used to solve the following three robust optimisation problem.
\begin{problem}\label{problem1}
Given the probabilistic models $f_1({\bf x})$ and $f_2({\bf x})$,
\begin{equation}\label{eq_constraint1}
\begin{aligned}
& \underset{{\bf x}\in\Omega_0}{\text{minimize}}
& & \mu_{{\omega}({\bf x})} =: \omega\mu_{{f}_1({\bf x})}^{pce} + (1-\omega)\mu_{{f}_2({\bf x})}^{pce}, \\
& \text{subject to}
& & \sigma_{f_1\left({\bf x}\right)}^{pce} < \sigma_1, \quad \sigma_{f_2\left({\bf x}\right)}^{pce} < \sigma_2.
\end{aligned}
\end{equation}
with $\omega\in[0,1]$.
\end{problem}
The Pareto curve ${\bf p(\omega)}=\left(\mu_{f_1({\bf x}^*(\omega))}^{pce}, \mu_{f_2({\bf x}^*(\omega))}^{pce}\right)$ with ${\bf x}^*(\omega)=\underset{{\bf x}\in\Omega_0}{\text{argmin}} \mu_{{\omega}({\bf x})}$ is plotted in Figure \ref{pareto_constraint_1} (a) for four different constrained cases and one unconstrained case. First, we observe that the Pareto curve is pushed away from the origin as the constraints equally become more stringent, noting also this has a greater influence on the second objective $\mu_{f_2}^{pce}$. Consequently, the corresponding optimal design space is pushed away form the boundaries of the original design space $\Omega_0$ as shown in Figure \ref{pareto_constraint_1}(b). Secondly, the marked points in Figure \ref{pareto_constraint_1} indicate a compromise minimisation between the two objectives $\mu_{f_1}^{pce}$ and $\mu_{f_2}^{pce}$, where both objectives achieve a minimum of around $0.2$ ($328.096 Pa$ for pressure drop and $13.554 ^{\circ} C$ for the temperature deviation in the physical space) with corresponding $\omega=0.45$ in (\ref{eq_constraint1}). The optimal design point $\left(0, 0.4484\right)$ as shown in Figure \ref{pareto_constraint_1}(b) has $95\%$ confidence intervals $\left(0.166, 0.237\right)=0.2015\pm0.071$ for $f_1$ and $\left(0.158, 0.247\right)=0.2025\pm0.089$ for $f_2$ as indicated by the marked points in Figure \ref{response_surface_confidence_projection}. Thirdly, if want to shrink the confidence intervals so that our prediction is more robust, such as with a $95\%$ confidence interval of $\pm0.04$ (standard deviation of 0.02) for both objectives, we can move from point $\left(0.1991, 0.2019\right)$ on the blue curve in Figure \ref{pareto_constraint_1}(a) to the green curve with the same $\mu_{f_1}^{pce}$. In this case, we are more confident that we would be able to achieve that objective, which is now around $0.2$ for $\mu_{f_1}^{pce}$ (unchanged) and $0.3$ for $\mu_{f_2}^{pce}$. We briefly summarise as follows: the more we can reduce $\mu_{f_1}^{pce}$ ($\mu_{f_2}^{pce}$), the less we would reduce $\mu_{f_2}^{pce}$ ($\mu_{f_1}^{pce}$) -- moving from one end to other on each curve in Figure \ref{pareto_constraint_1} (a); the more confident we are in the reduction of $\mu_{f_1}^{pce}$ (moving from the blue curve across the green one, and towards the purple curve at a fixed $\mu_{f_1}^{pce}$ in Figure \ref{pareto_constraint_1}(a)), the less confident we are in the reduction in $\mu_{f_2}^{pce}$. We can achieve this by keeping the second design variable constant and increasing only the first design variable as shown in \ref{pareto_constraint_1} (b). The reason we can do this is because the response surface for $f_1$ is almost constant for a fixed $x_2$ as shown in Figure \ref{response_surface_confidence}.

\begin{figure}[h!]
\begin{minipage}[t]{0.5\linewidth}
\centering  
\includegraphics[width=2.3in,angle=0]{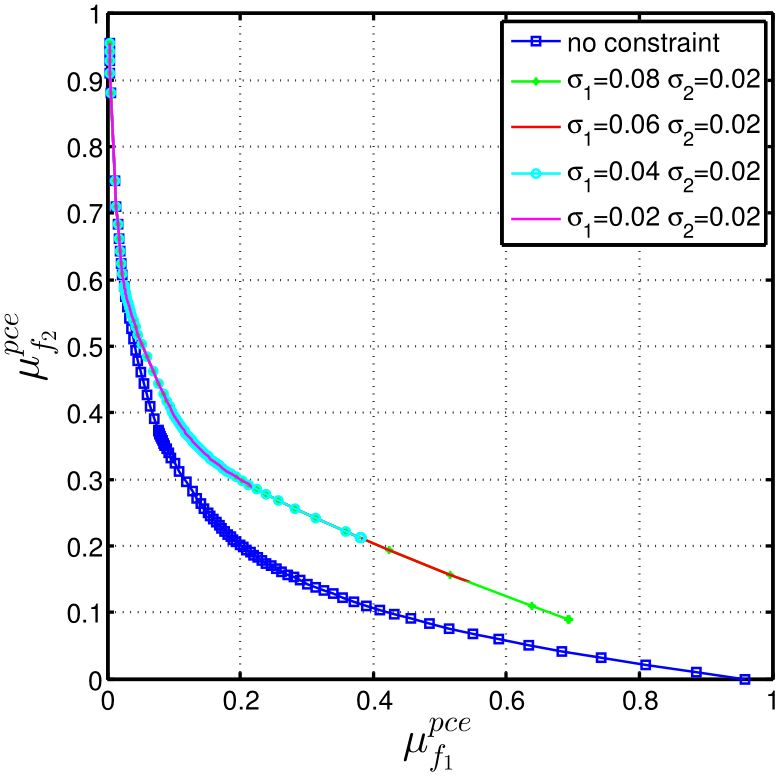}
\captionsetup{justification=centering}
\caption*{\scriptsize(c) Pareto curve.}
\end{minipage}
\begin{minipage}[t]{0.5\linewidth}
\centering  
\includegraphics[width=2.3in,angle=0]{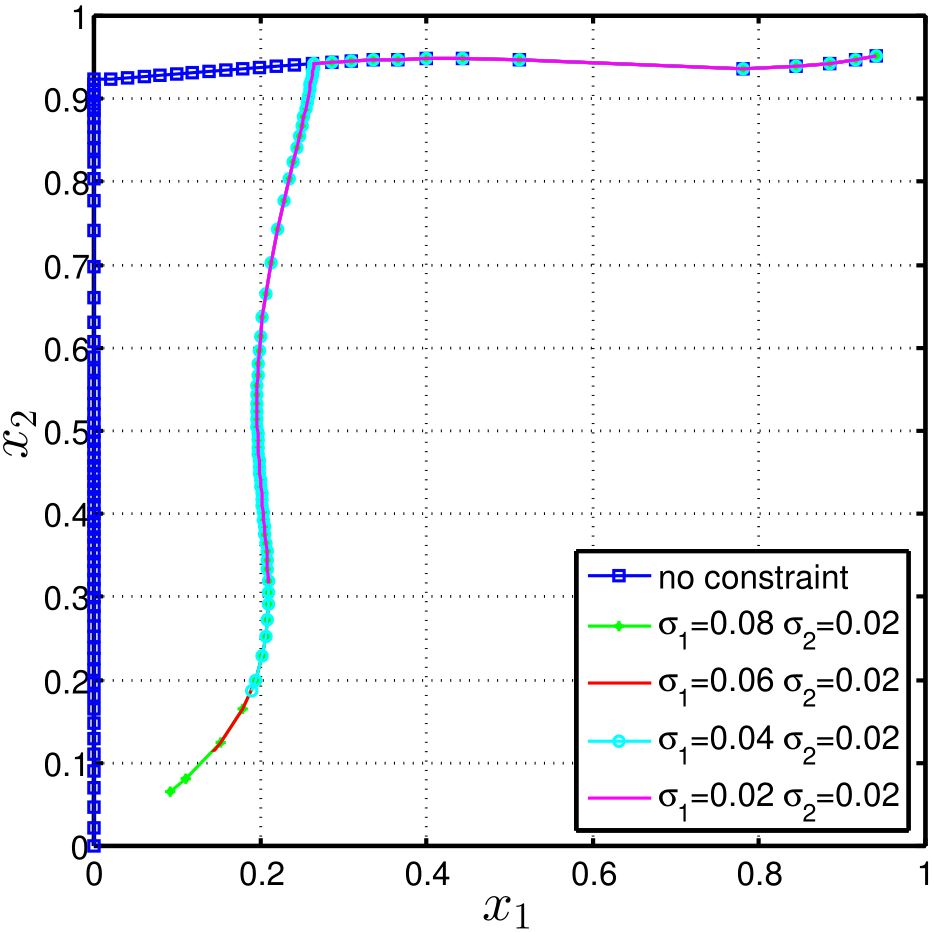}
\captionsetup{justification=centering}
\caption*{\scriptsize(d) Opimal design space.}
\end{minipage}
\caption {\scriptsize Pareto curve  ${\bf p(\omega)}$ and the corresponding designs for different cases of constraints.} 
\label{pareto_constraint_2}
\end{figure}
\begin{figure}[h!]
\centering  
\includegraphics[width=3in,angle=0]{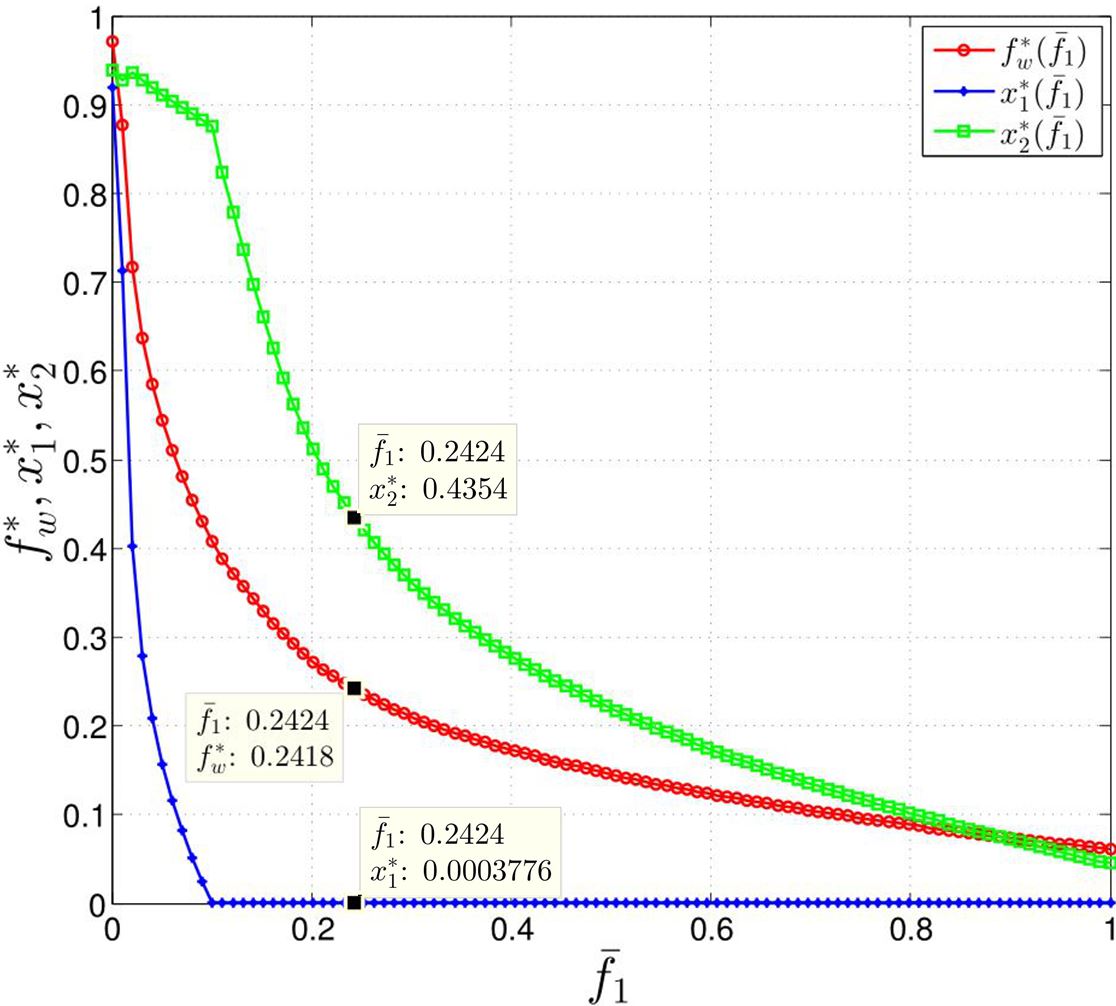}
\captionsetup{justification=centering}
\caption {\scriptsize Solution of Problem \ref{problem2} as a function of upper bound $\bar{f_1}$.} 
\label{fwp2}
\end{figure}

We test another case of constraints in Figure \ref{pareto_constraint_2}, where the upper bound of $\sigma_{f_1}^{pce}$ varies while the upper bound of $\sigma_{f_2}^{pce}$ stays the same. It can be seen from Figure \ref{pareto_constraint_2} that the first parameter $x_1$ of the optimal designs is almost unchanged while the second one $x_2$ varies rapidly as the upper bound of $\sigma_1$ varies. This is different in the first test case shown in Figure \ref{pareto_constraint_1} where the main variation of the optimal designs lies in the first parameter $x_1$.
\begin{problem}\label{problem2}
Given the probabilistic models $f_1({\bf x})$ and $f_2({\bf x})$,
\begin{equation}
\begin{aligned}
& \underset{{\bf x}\in\Omega_0}{\text{minimize}}
& & f_w({\bf x}) =: \mu_{{f}_2({\bf x})} + 2\sigma_{{f}_2({\bf x})}, \\
& \text{subject to}
& & \mu_{f_1\left({\bf x}\right)} + 2\sigma_{f_1\left({\bf x}\right)} < \bar{f_1}.
\end{aligned}
\end{equation}
\end{problem}
The objective function $f_w({\bf x})$ in Problem \ref{problem2} defines the``worst case" for $f_2$ with 95\% confidence, and a minimisation of $f_w({\bf x})$ given a safe (95\% confidence) upper bound $\bar{f_1}$ for $f_1$. We plot, in Figure \ref{fwp2}, $f_w^*=\underset{{\bf x}\in\Omega_0}{\text{min}} {f_{\omega}(x_1,x_2)}$ and $\left(x_1^*, x_2^*\right)=\underset{{\bf x}\in\Omega_0}{\text{argmin}} {f_{\omega}(x_1,x_2)}$ as functions of $\bar{f_1}$, i.e.: $f_w^*\left(\bar{f_1}\right)$, $x_1^*\left(\bar{f_1}\right)$ and $x_2^*\left(\bar{f_1}\right)$. It can be seen from Figure \ref{fwp2} that minimising $f_w$ is always accompanied with an increasing upper bound $\bar{f_1}$. The meaning of the marked points (which correspond to the same points in the design space) shown in Figure \ref{fwp2} is: we can reduce $f_w$ to $f_w^*=0.2418$ by choosing $x_1=x_1^*=0$ and $x_2=x_2^*=0.4354$, at the same time we also have $95\%$ confidence that $f_1<0.2424$. 
\begin{problem}\label{problem3}
Given the probabilistic models $f_1({\bf x})$ and $f_2({\bf x})$,
\begin{equation}
\begin{aligned}
& \underset{{\bf x}\in\Omega_0}{\text{minimize}}
& & f_w({\bf x}) =: \mu_{{f}_1({\bf x})} + 2\sigma_{{f}_1({\bf x})}, \\
& \text{subject to}
& & \mu_{f_2\left({\bf x}\right)} + 2\sigma_{f_2\left({\bf x}\right)} < \bar{f_2}.
\end{aligned}
\end{equation}
\end{problem}

\begin{figure}[h!]
\centering  
\includegraphics[width=3in,angle=0]{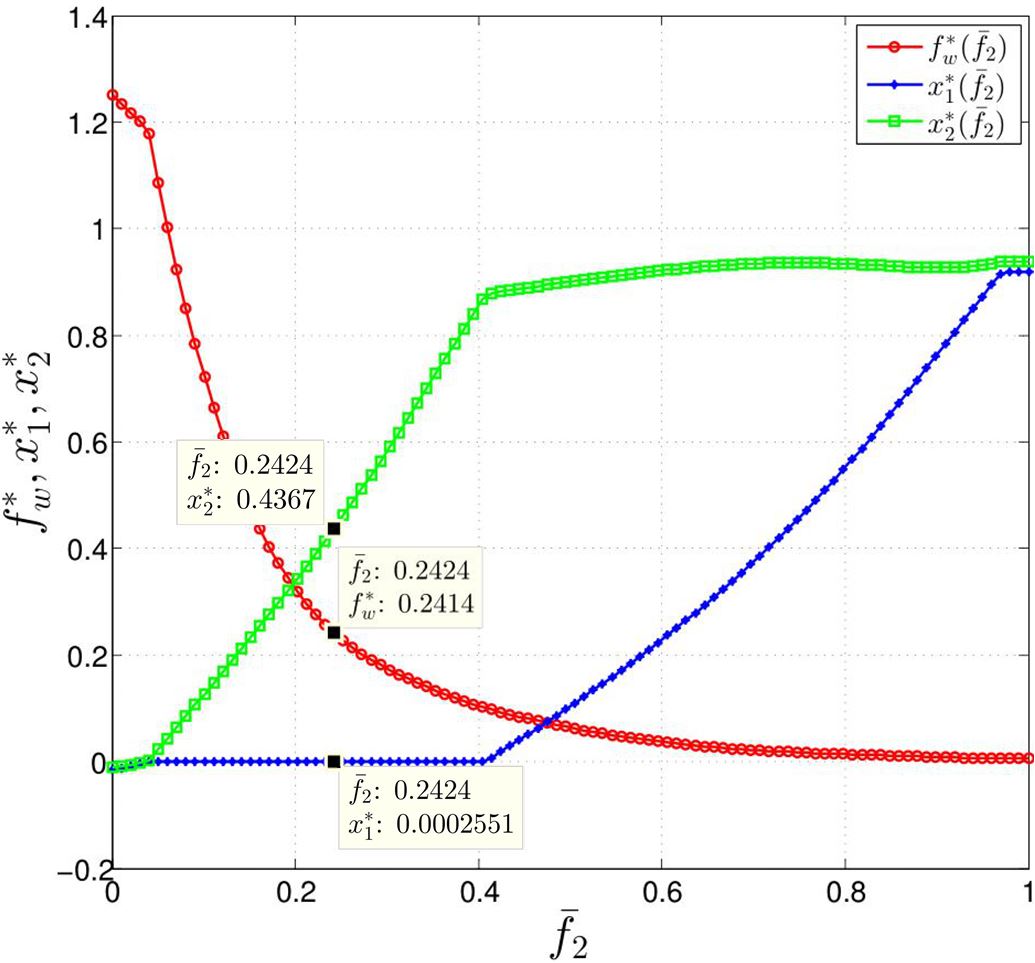}
\captionsetup{justification=centering}
\caption {\scriptsize Solution of Problem \ref{problem3} as a function of upper bound $\bar{f_2}$.} 
\label{fwp3}
\end{figure}
Similar to Problem \ref{problem2}, Problem \ref{problem3} can be interpreted as minimising the ``worst case" of $f_1$ given a stringent (95\% guaranteed) constraint for $f_2$. It can be seen from Figure \ref{fwp3} that we can reduce $f_1$, with 95\% confidence, to $f_w^*=0.2414$ by choosing $x_1=x_1^*=0$ and $x_2=x_2^*=0.4367$, at the same time we also have 95\% confidence that $f_2<0.2424$. This is consistent with the result obtained by solving Problem \ref{problem2}.
\section{Conclusion}\label{sec_conclusion}
PCR systems provide effective methods for rapid diagnosis of infectious diseases and are playing a vital role in the public health response to COVID-19. Thermal flows in PCR systems presents complex, multi-objective optimisation problems which need to account for uncertainties caused by manufacturing errors as well as variations in flow rate and heat flux. This paper demonstrates that combining accurate conjugate heat transfer simulations with Gaussian Process Regression (GPR) and Polynomial Chaos Expansions (PCE) can create an efficient probabilistic surrogate model for the robust optimisation of CFPCR systems with respect to temperature uniformity and pressure drop. The methodology can be extended to incorporate other important PCR objectives such as maximising the DNA amplification efficiency or minimising heating power requirements. This approach is relatively easy to implement using the existing free libraries {\bf GPy}, {\bf ChaosPy} and {\bf SciPy}.

The three different robust design optimisation formulations considered result in a series of Pareto fronts of non-dominated solutions which vary depending on the relative importance of the standard deviation to mean performance. These provide a convenient means of balancing competing objectives and demonstrate how the inclusion of robustness requirements, where the standard deviation and mean of objectives must be accounted for simultaneously, leads to increased design conservatism and reductions in performance compared to deterministic design optima. This study will be extended to consider other sources on uncertainty in PCR design, for example due to flow and thermal operating conditions.

\section*{Conflicts of interest}
On behalf of all authors, the corresponding author states that there is no conflict of interest.

\section*{Replication of results}
All the Python code of implementing the numerical tests in this paper are attached to the following appendices of this paper.

\appendix
\section{Python code for testing the convergence of PCE method and its validation}\label{appendix_pce_convergence}
In this appendix, we present the Python code for testing the PCE method's convergence in terms of the order of the orthogonal polynomial basis. We also validate the PCE by combination of two normal distributions, and compare its efficiency against the Monte Carlo method.
\begin{lstlisting}[language=Python]
import numpy as np 
import chaospy as cp
import pandas as pd 
import GPy
###################
dim_in=2
dim_out=2
dim=dim_in+dim_out
###################
pd.set_option('precision',16)
data = pd.read_csv('train.txt',header=None) 

x = data.iloc[:, 0:dim_in].values 
y = data.iloc[:, dim_in:dim].values 

rbf = GPy.kern.RBF(input_dim=dim_in, variance=1, lengthscale=1)
gp = GPy.models.GPRegression(x, y, rbf)

gp.optimize()
gp = GPy.models.GPRegression(x, y, rbf,noise_var=1.e-10)

c0 = cp.Normal(0.5, 0.025)
c1 = cp.Normal(0.6, 0.025)

distribution = cp.J(c0,c1)

def pce(poly_order):
    nodes,weights = cp.generate_quadrature(poly_order,
    distribution,rule="Gaussian")      

    x=np.transpose(nodes)
    y_pred, var = gp.predict(x)
    
    f1=y_pred[:,0]
    f2=y_pred[:,1]
    '''
    sd=np.sqrt(var)
    f1=y_pred[:,0]
    f2=y_pred[:,1]
    for i in np.arange(np.size(f1)):
        noise = np.random.normal(0, sd[i,0], 1)
        f1[i] += noise
        f2[i] += noise
    '''
    polynomials = cp.orth_ttr(order=poly_order,dist=distribution)

    model_approx = cp.fit_quadrature(polynomials,nodes,weights,f1)
   #model_approx = cp.fit_quadrature(polynomials,nodes,weights,f2)
                                
    mean = cp.E(model_approx, distribution)
    deviation = cp.Std(model_approx, distribution)

    print("mean, deviation=",mean,deviation)

    file=open("pce_convergence.txt","a")
    file.write("%.16f  %.16f\n" % (mean,deviation))
    file.close()

for i in np.arange(6):
    poly_order=i+1
    pce(poly_order)
\end{lstlisting}

\subsection{Linear combination of two normal distributions}
We first test the PCE code using $aX_1+bX_2\sim \mathcal{N}(a\mu_1+b\mu_2, (a\sigma_1)^2+(b\sigma_2)^2)$, given $X_1\sim \mathcal{N}(\mu_1, \sigma_1^2)$ and $X_2\sim \mathcal{N}(\mu_2, \sigma_2^2)$. Since this is a linear relation, the PCE code exactly computes the mean and standard deviation, using a first order basis. For example, $a=b=1$, $\mu_1=0.5$, $\mu_2=0.7$, $\sigma_1=3$ and $\sigma_2=4$, the PCE model gives exact $E(X_1+X_2)=1.2$ and $\sigma(X_1+X_2)=5$ using a first order (or higher) basis,
or $a=1$, $b=2$, $\mu_1=0.3$, $\mu_2=0.5$, $\sigma_1=0.3$ and $\sigma_2=0.2$, the PCE model gives exact $E(X_1+2X_2)=1.3$ and $\sigma(X_1+2X_2)=0.5$.

\begin{figure}[h!]
	\centering  
	\includegraphics[width=2.3in,angle=0]{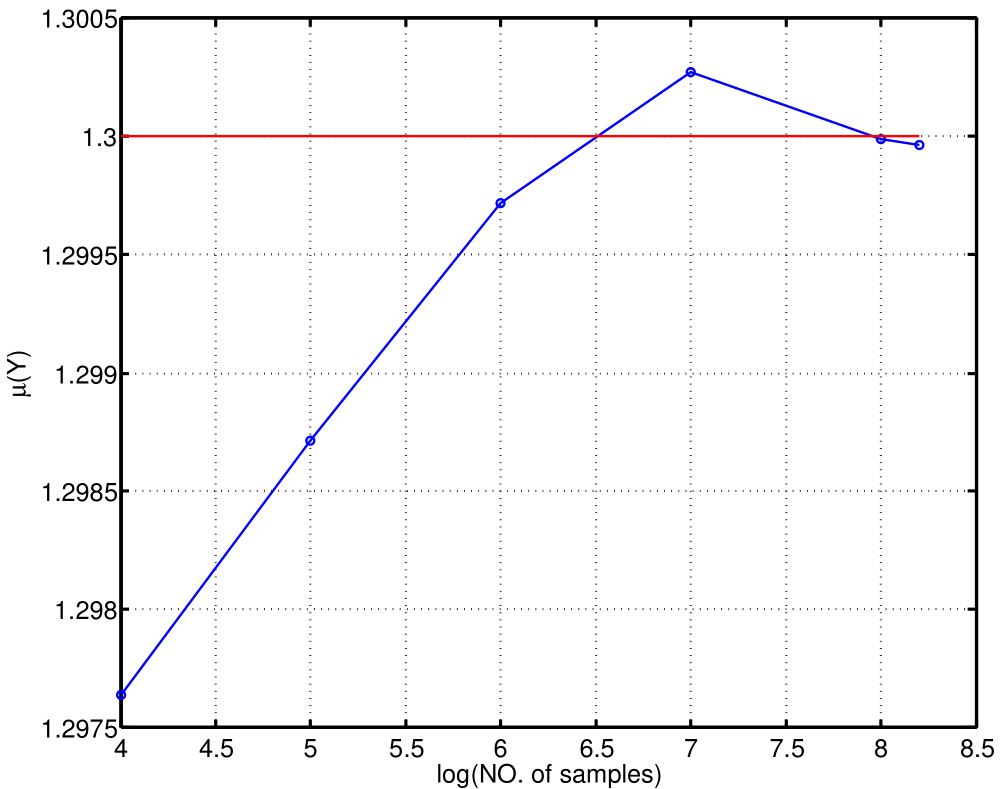}
	\includegraphics[width=2.3in,angle=0]{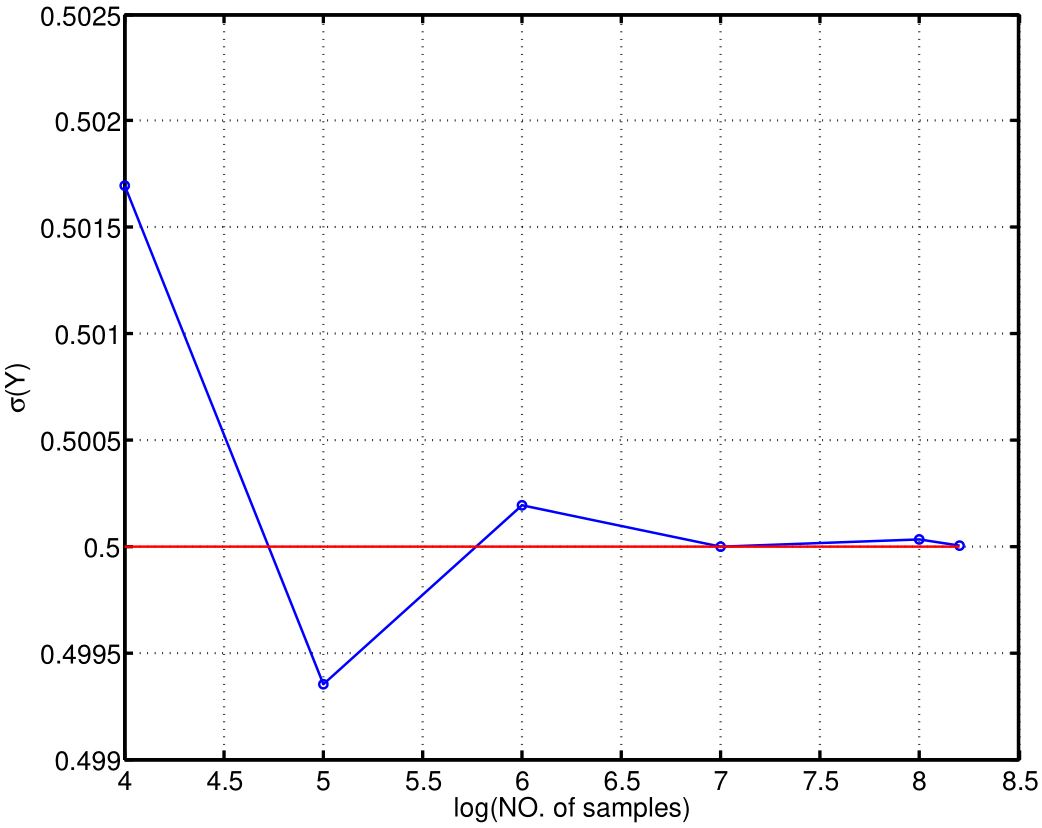}	
	\captionsetup{justification=centering}
	\caption {\scriptsize Convergence of the mean (left) and standard deviation (right) as a function of the number of sampling points.} 
	\label{convergence_sampling}
\end{figure}

If we use the Monte Carlo method to compute the statistics of $X_1+2X_2$, a random test gives: $\mu=1.2993$ and $\sigma=0.5013$ using $10^5$ samples, $\mu=1.3007$ and $\sigma=0.4965$ using $10^6$ samples, and $\mu=1.3282$ and $\sigma=0.5002$ using $10^7$ samples. A convergence of this sampling is shown in Figure \ref{convergence_sampling}, from which we can see how poorly the Monte Carlo method converges: one needs a very large number of the sampling points in order to gain an accurate approximation.

\subsection{An example of non-linear function}
\begin{figure}[h!]
	\centering  
	\includegraphics[width=2.3in,angle=0]{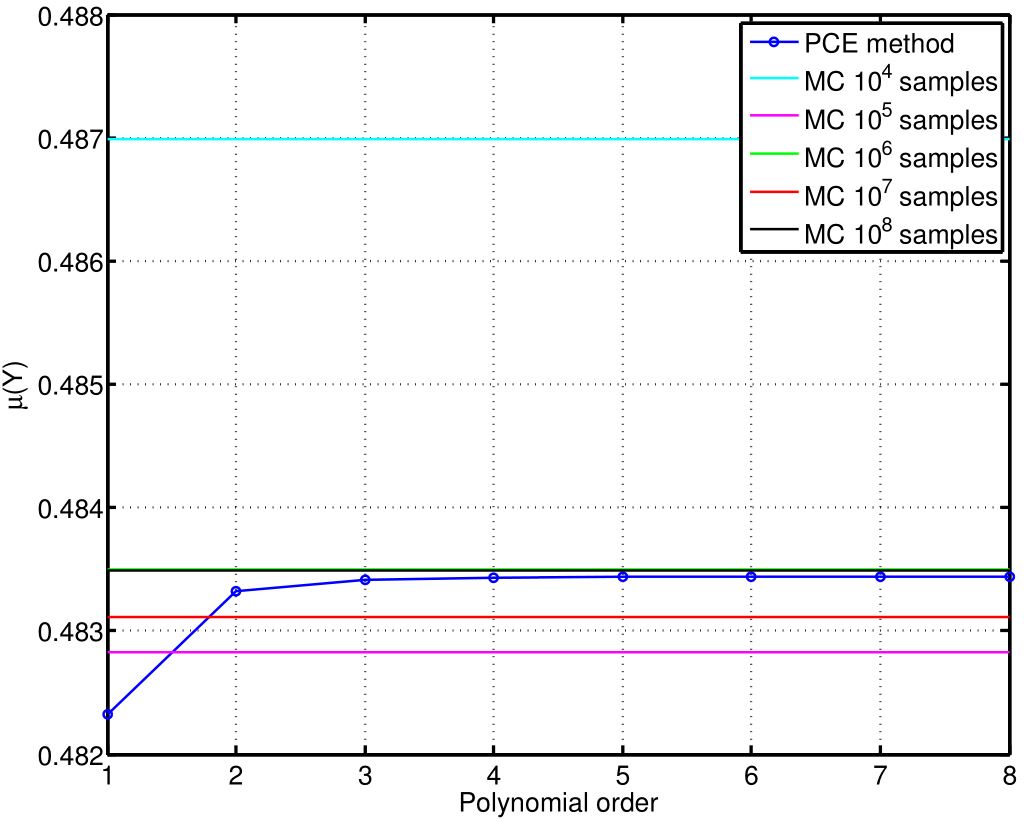}
	\includegraphics[width=2.3in,angle=0]{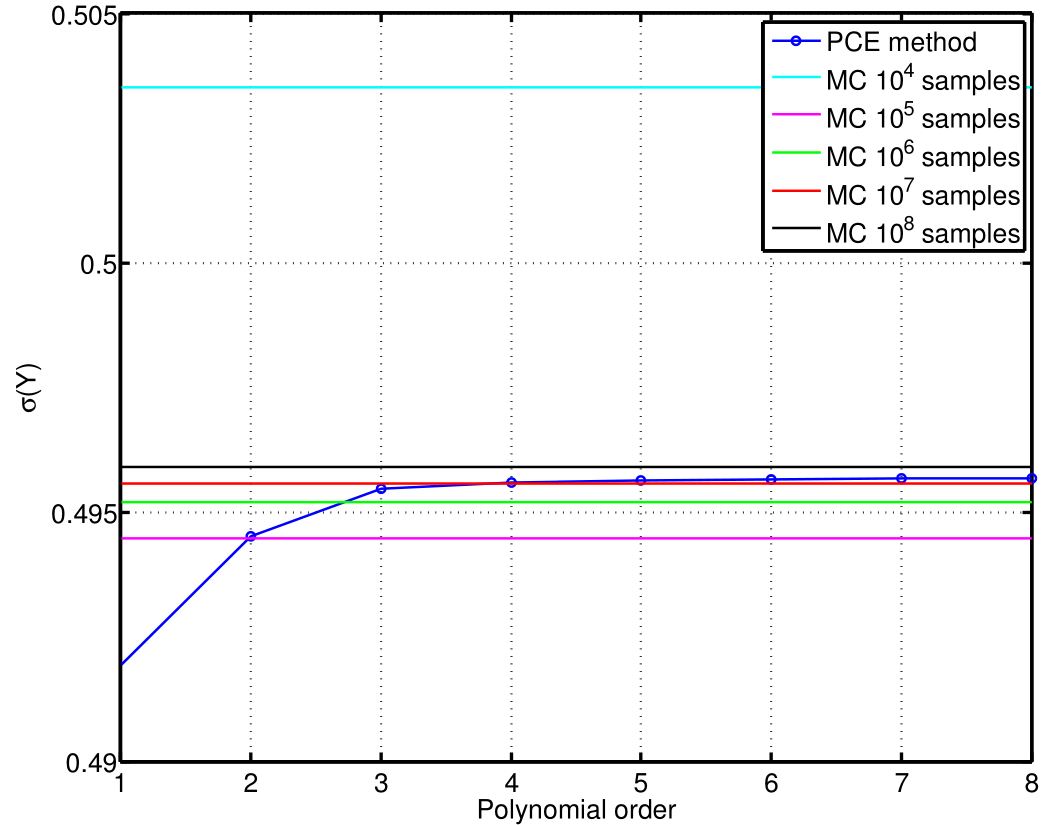}	
	\captionsetup{justification=centering}
	\caption {\scriptsize Convergence of the mean (left) and standard deviation (right) as a function of the polynomial order of the PCE method.} 
	\label{convergence_pce}
\end{figure}

We generate data using the following example function:
\begin{equation}
	Y=sin(X_1)/cos(X_2)+X_2X_2,
\end{equation}
with $X_1\sim \mathcal{N}(\mu_1, \sigma_1^2)$ and $X_2\sim \mathcal{N}(\mu_2, \sigma_2^2)$. For $\mu_1=0.3$, $\mu_2=0.5$, $\sigma_1=0.3$ and $\sigma_2=0.2$, the convergence of the PCE method together with the convergence of Monte Carlo method are shown in Figure \ref{convergence_pce}, from which it can be seen that the PCE is much cheaper.

One should notice that assuming a Gaussian input (or other distribution) is the prerequisite for using the PCE method, while Monte Carlo method needs no assumption of the input variables. This is the essential reason why the PCE method is more efficient than the Monte Carlo method.

\section{Python code to predict on uniform grids using PCE method}\label{appendix_pce_predict}
\begin{lstlisting}[language=Python]
import numpy as np 
import chaospy as cp
import pandas as pd 
import GPy
###################
dim_in=2
dim_out=2
dim=dim_in+dim_out
###################
pd.set_option('precision',16)
data = pd.read_csv('train.txt',header=None) 

x = data.iloc[:, 0:dim_in].values 
y = data.iloc[:, dim_in:dim].values 

rbf = GPy.kern.RBF(input_dim=dim_in, variance=1, lengthscale=1)
gp = GPy.models.GPRegression(x, y, rbf)

gp.optimize()
gp = GPy.models.GPRegression(x, y, rbf,noise_var=1.e-10)

poly_order=4
def pce(distribution):
    nodes,weights = cp.generate_quadrature(poly_order,
    distribution,rule="Gaussian")      

    y_pred, var = gp.predict(np.transpose(nodes))
    
    f1=y_pred[:,0]
    f2=y_pred[:,1]

    polynomials = cp.orth_ttr(order=poly_order,dist=distribution)

    model_approx = cp.fit_quadrature(polynomials,nodes,weights,f1)
   #model_approx = cp.fit_quadrature(polynomials,nodes,weights,f2)
                                
    mean = cp.E(model_approx, distribution)
    deviation = cp.Std(model_approx, distribution)

    print("mean, deviation=",mean,deviation)

    file=open("pce_predict.txt","a")
    file.write("%.16f  %.16f\n" % (mean,deviation))
    file.close()

xset = np.linspace(0, 1, 100)
yset = np.linspace(0, 1, 100)
for xm in xset:
    for ym in yset:
        c0 = cp.Normal(xm, 0.025)
        c1 = cp.Normal(ym, 0.025)
        distribution = cp.J(c0,c1)
        pce(distribution)
\end{lstlisting}
\section{Python code for optimisation based up the PCE model}\label{appendix_pce_optimisation}
We only provide in this section the Python code for solving Problem \ref{problem1}, which is also a template for Problems \ref{problem2} and \ref{problem3}.
\begin{lstlisting}[language=Python]
import numpy as np 
import pandas as pd 
import GPy
from scipy.optimize import minimize
###################
dim_in=2
dim_out=4
dim=dim_in+dim_out
###################
def gpr(x):
	x=x.reshape(1,len(x))
	yp = gp.predict(x)[0]
	return yp[0,0]*omega+yp[0,1]*(1.-omega)
def gpr_mean(x):
	x=x.reshape(1,len(x))
	yp = gp.predict(x)[0]
	return [yp[0,0],yp[0,1]]
def gpr_sigma(x):
	x=x.reshape(1,len(x))
	yp = gp.predict(x)[0]
	return [yp[0,2],yp[0,3]]

pd.set_option('precision',16)
data = pd.read_csv('pce_predict.txt',header=None) 

x_train = data.iloc[:, 0:dim_in].values 
y_train = data.iloc[:, dim_in:dim].values 

rbf = GPy.kern.RBF(input_dim=dim_in, variance=1, lengthscale=1)
gp = GPy.models.GPRegression(x_train, y_train, rbf)

'''Run optimization'''
gp.optimize()

'''Obtain optimized kernel parameters'''
noise=gp.Gaussian_noise.variance
print("optimized noise: \n",noise[0])
l = gp.rbf.lengthscale.values
sigma_f = np.sqrt(gp.rbf.variance.values)
print("optimized kernel parameters: \n",l,sigma_f)

rbf = GPy.kern.RBF(input_dim=dim_in, variance=sigma_f**2,
 lengthscale=l)
'''Fix the noise variance to known value'''
gp.Gaussian_noise.variance = noise
gp.Gaussian_noise.variance.fix()
gp = GPy.models.GPRegression(x_train, y_train, rbf)

x0=[0.5,0.5]
bb=((0,1),(0,1))
sigma1=0.08
sigma2=0.02

from scipy.optimize import NonlinearConstraint
nonlinear_constraint = NonlinearConstraint(gpr_sigma,
[-np.inf,-np.inf], [sigma1,sigma2])

OmegaSet = np.linspace(0, 1, 100)
f=open("pareto.txt","w+")
for omega in OmegaSet:
    #res=minimize(gpr, x0, bounds=bb})
    res = minimize(gpr, x0, method='trust-constr', 
    constraints=[nonlinear_constraint],
    options={'xtol': 1e-12,'verbose': 1}, bounds=bb)      
    print('minimum:', omega, res.x, res.fun,'\n')
    mean=gpr_mean(res.x)
    f.write("%.10f %.10f %.10f %.10f\n" % 
    (res.x[0],res.x[1],mean[0],mean[1]))

f.close()
\end{lstlisting}


%
%

\bibliographystyle{spmpsci}      
\bibliography{mybibfile}   

%
%

\end{document}